\documentclass[aps,pra,twocolumn,showpacs,floatfix,longbibliography]{revtex4-2}
\usepackage{graphicx,amsfonts,amssymb,amsmath}
\usepackage{textcomp} 
\usepackage{esint}
\usepackage[english]{babel}
\usepackage{afterpage}
\usepackage{bbold}
\usepackage{soul}
\usepackage[colorlinks,linktocpage,citecolor=blue,linkcolor=red,urlcolor=blue]{hyperref}
\usepackage{xcolor}
\usepackage{booktabs}

\def\xiloc{\xi_{\rm loc}} 
\def\kdis{k_{\rm dis}}
\def\kmax{k_{\rm max}}

\begin{document}
	
\graphicspath{{./figures/}}
	
\allowdisplaybreaks
	

%


\newcommand{\oldnew}[2]{\marginpar{\scriptsize \textcolor{red}{correction}}{\textcolor{red}{#2}}}
\newcommand{\suppressed}[1]{\marginpar{\scriptsize \textcolor{red}{correction}}{\textcolor{red}{\st{#1}}}}
\newcommand{\correction}[1]{\marginpar{\textcolor{red}{\scriptsize #1}}}
\newcommand{\marge}[1]{\marginpar{\scriptsize #1}}
\newcommand{\remarque}[1]{\marginpar{\scriptsize Remarque}{\it [#1]}}

%

\newcommand\blfootnote[1]{
	\begingroup
	\renewcommand\thefootnote{}\footnote{#1}
	\addtocounter{footnote}{-1}
	\endgroup
}

\def\rhoeq{\hat\rho_{\rm eq}}

\newcommand{\new}[1]{{\bf #1}}
\newlength{\textlarg}
\newcommand{\redbar}[1]{\textcolor{red}{\st{#1}}} 
\newcommand{\bluebar}[1]{\textcolor{blue}{\st{#1}}} 

\newcommand{\normord}[1]{:\mathrel{#1}:}

\newcommand{\beq}{\begin{equation}}
\newcommand{\eeq}{\end{equation}}
\newcommand{\bfig}{\begin{figure}}
\newcommand{\efig}{\end{figure}}
\newcommand{\bline}{\begin{multline}}
\newcommand{\eline}{\end{multline}}
\newcommand{\bremark}{\begin{quotation} \noindent \small }
\newcommand{\eremark}{\end{quotation}}
\newcommand{\llbrace}{\left\lbrace}  
\newcommand{\rrbrace}{\right\rbrace}
\newcommand{\lbraket}{\left[}
\newcommand{\rbraket}{\right]}
\newcommand{\llangle}{\left\langle}
\newcommand{\rrangle}{\right\rangle} 

\newcommand{\Tr}{{\rm Tr}} 
\newcommand{\tr}{{\rm tr}} 
\newcommand{\sgn}{\,{\rm sgn}} 
\newcommand{\mean}[1]{\langle #1 \rangle}
\newcommand{\commu}[2]{[#1,#2]} 
\newcommand{\bra}[1]{\langle#1|}
\newcommand{\ket}[1]{|#1\rangle}
\newcommand{\braket}[2]{\langle #1|#2\rangle}
\newcommand{\ketbra}[2]{|#1\rangle\langle#2|}
\newcommand{\dbraket}[3]{\langle #1|#2|#3\rangle}
\newcommand{\tens}[1]{\overleftrightarrow{#1}}  
\newcommand{\vac}{|{\rm vac}\rangle} 
\newcommand{\bravac}{\langle{\rm vac}|}
\newcommand{\const}{{\rm const}} 
\newcommand{\unif}{{\rm unif.}} 
\newcommand{\atanh}{\,{\rm atanh}}
\newcommand{\cotanh}{\,{\rm cotanh}}

\newcommand{\ie}{i.e.\xspace}
\newcommand{\iet}{i.e.}
\newcommand{\eg}{e.g.\xspace}
\newcommand{\cc}{{\rm c.c.}} 
\newcommand{\hc}{{\rm H.c.}} 
\newcommand{\etal}{{\it et al. }}
\newcommand\eme{$^{\mbox{\footnotesize ème}}$\xspace}

\newcommand{\jhatbf}{\hat {\textbf \jold}} 
\newcommand{\Jhatbf}{\hat {\textbf \J}} 
\newcommand{\jhat}{\hat {\jmath}} 
\newcommand{\Jhat}{\hat {J}} 
\newcommand{\jbf}{\textbf j}
\newcommand{\Jbf}{\textbf J}

\def\chibf{\boldsymbol{\chi}}
\def\down{\downarrow}
\def\eps{\epsilon}
\def\gam{\gamma} 
\def\alphabf{\boldsymbol{\alpha}}
\def\phibf{\boldsymbol{\phi}}
\def\varphibf{\boldsymbol{\varphi}}
\def\varphibfs{\boldsymbol{\varphi}_<}
\def\varphibfl{\boldsymbol{\varphi}_>}
\def\varphis{\varphi_{<}}
\def\varphil{\varphi_{>}}
\def\psibf{\boldsymbol{\psi}}
\def\thetabf{\boldsymbol{\theta}}
\def\Ome{\Omega}
\def\omeD{{\omega_D}} 
\def\bfOme{\boldsymbol{\Omega}} 
\def\Omebf{\boldsymbol{\Omega}} 
\def\lamb{\lambda}
\def\Lamb{\Lambda}
\def\sig{\sigma}
\def\Sig{\Sigma}
\def\sigp{{\sigma'}} 
\def\bfsig{\boldsymbol{\sigma}} 
\def\sigbf{\boldsymbol{\sigma}} 
\def\bfSig{\boldsymbol{\Sigma}} 
\def\The{\Theta} 
\def\up{\uparrow}

\def\epsk{\epsilon_{\bf k}} 
\def\epsp{\epsilon_{\bf p}} 
\def\xik{\xi_{\bf k}} 
\def\txik{\tilde\xi_{\bf k}} 
\def\xip{\xi_{\bf p}} 
\def\epsq{\epsilon_{\bf q}} 
\def\xiq{\xi_{\bf q}} 
\def\xikq{\xi_{{\bf k}+{\bf q}}} 
\def\Ek{E_{\bf k}} 
\def\Ep{E_{\bf p}}
\def\Eq{E_{\bf q}}
\def\Heff{\hat H_{\rm eff}}
\def\Hem{\hat H_{\rm em}}
\def\Hint{\hat H_{\rm int}}
\def\Hloc{\hat H_{\rm loc}}
\def\HMF{\hat H_{\rm MF}}
\def\HLL{\hat H_{\rm LL}}
\def\Hdis{\hat H_{\rm dis}}
\def\Sem{S_{\rm em}}
\def\SMF{S_{\rm MF}} 
\def\SHF{S_{\rm HF}} 
\def\SRPA{S_{\rm RPA}} 
\def\Sint{S_{\rm int}} 
\def\Sloc{S_{\rm loc}}
\def\TN{T_{\rm N}} 
\def\TNHF{T^{\rm HF}_{\rm N}} 
\def\Zloc{Z_{\rm loc}} 
\def\ZMF{Z_{\rm MF}} 
\def\ZHF{Z_{\rm HF}} 
\def\ZRPA{Z_{\rm RPA}} 
\def\RPA{{\rm RPA}}
\def\loc{{\rm loc}} 
\def\pp{{\rm pp}}
\def\ph{{\rm ph}} 
\def\ch{{\rm ch}}
\def\sp{{\rm sp}} 
\def\qtf{q_{\rm TF}}
\def\epstf{\eps^{}_{\rm TF}} 
\def\epsrpa{\eps^{}_{\rm RPA}} 
\def\chinnzpp{\chi_{nn}^{0}{}\!\!\!''}
\def\SigHF{\Sigma_{\rm HF}}
\def\psicl{\psi_{\rm cl}} 

\def\half{\frac{1}{2}}
\def\dhalf{\dfrac{1}{2}}
\def\third{\frac{1}{3}} 
\def\quarter{\frac{1}{4}}

\def\qr{{\bf q}\cdot{\bf r}}
\def\wt{\omega t} 

\def\a{{\bf a}}
\def\b{{\bf b}}
\newcommand{\cv}{{\bf c}} 
\def\e{{\bf e}}
\def\f{{\bf f}}
\def\g{{\bf g}}
\def\h{{\bf h}}
\def\jold{\char"11}
\def\j{{\bf j}}
\def\k{{\bf k}}
\def\l{{\bf l}}
\def\ellbf{\bm{\ell}} 
\def\m{{\bf m}}
\def\n{{\bf n}} 
\def\p{{\bf p}} 
\def\q{{\bf q}}
\def\r{{\bf r}}
\def\t{{\bf t}}
\def\u{{\bf u}}
\newcommand{\vv}{{\bf v}}
\def\x{{\bf x}}
\def\y{{\bf y}} 
\def\z{{\bf z}} 
\def\A{{\bf A}}
\def\B{{\bf B}}
\def\D{{\bf D}} 
\def\E{{\bf E}} 
\def\F{{\bf F}} 
\def\H{{\bf H}}  
\def\J{{\bf J}}
\def\K{{\bf K}} 

\def\G{{\bf G}}
\def\L{{\bf L}}
\def\M{{\bf M}}  
\def\O{{\bf O}} 
\def\P{{\bf P}} 
\def\Q{{\bf Q}} 
\def\R{{\bf R}}
\def\S{{\bf S}}
\def\U{{\bf U}} 
\def\V{{\bf V}} 
\def\X{{\bf X}} 
\def\Y{{\bf Y}} 
\def\epsbf{\boldsymbol{\epsilon}}
\def\betabf{\boldsymbol{\beta}}
\def\deltabf{\boldsymbol{\delta}}
\def\mubf{\boldsymbol{\mu}}
\def\nablabf{\boldsymbol{\nabla}}
\def\rhobf{\boldsymbol{\rho}}
\def\sigmabf{\boldsymbol{\sigma}} 
\def\Pibf{\boldsymbol{\Pi}}
\def\pibf{\boldsymbol{\pi}}

\def\para{\parallel}
\def\kpara{{k_\parallel}}
\def\kperp{{k_\perp}} 
\def\kperpp{{k_\perp'}} 
\def\qperp{{q_\perp}} 
\def\tperp{{t_\perp}} 

\def\w{\omega}
\def\wn{\omega_n}
\def\wm{\omega_m}
\def\wnu{\omega_\nu}
\def\wp{\omega_p} 
\def\dmu{{\partial_\mu}}
\def\dnu{{\partial_\nu}}
\def\dl{{\partial_l}}  
\def\dt{\partial_t} 
\def\tdt{\tilde\partial_t}
\def\dk{\partial_k}
\def\tdk{\tilde\partial_k}
\def\dx{\partial_x}
\def\dxtilde{\partial_{\tilde x}}
\def\dy{\partial_y} 
\def\dw{\partial_{\w}}
\def\dtau{{\partial_\tau}} 
\def\dtautilde{{\partial_{\tilde\tau}}}   
\def\det{{\rm det}} 
\def\Pf{{\rm Pf}}
\def\diag{{\rm diag}}

\def\dsum{\displaystyle \sum}
\def\dint{\displaystyle \int} 
\def\intt{\int_{-\infty}^\infty dt} 
\def\inttp{\int_{-\infty}^\infty dt'} 
\def\intk{\int_{\bf k}} 
\def\intkd{\int \frac{d^dk}{(2\pi)^d}}
\def\intq{\int_{\bf q}} 
\def\intr{\int d^dr}  
\def\dintr{\displaystyle \int d^dr} 
\def\intrp{\int d^dr'}
\def\dinttau{\displaystyle \int_0^\beta d\tau}
\def\dinttaup{\displaystyle \int_0^\beta d\tau'}
\def\inttau{\int_0^\beta d\tau}
\def\inttaup{\int_0^\beta d\tau'}
\def\intx{\int d^{d+1}x} 
\def\inttaur{\int_0^\beta d\tau \int d^dr}
\def\intinf{\int_{-\infty}^\infty}
\def\dinttaur{\displaystyle \int_0^\beta d\tau \int d^dr}
\def\dintinf{\displaystyle \int_{-\infty}^\infty}
\def\intw{\int_{-\infty}^\infty \frac{d\w}{2\pi}}
\def\sumr{\sum_{\bf r}} 

\def\calA{{\cal A}}
\def\calAbf{\bm{{\cal A}}}
\def\calB{{\cal B}} 
\def\calC{{\cal C}} 
\def\dt{\partial_t}
\def\calD{{\cal D}}
\def\calE{{\cal E}}
\def\calF{{\cal F}} 
\def\calFbf{\bm{{\cal F}}}
\def\calG{{\cal G}}
\def\calH{{\cal H}}
\def\calI{{\cal I}}
\def\calJ{{\cal J}}
\def\calK{{\cal K}}
\def\calL{{\cal L}} 
\def\calM{{\cal M}} 
\def\calN{{\cal N}}
\def\calO{{\cal O}}
\def\calP{{\cal P}}  
\def\calR{{\cal R}} 
\def\calS{{\cal S}}
\def\calT{{\cal T}}
\def\calU{{\cal U}}
\def\calV{{\cal V}}
\def\calX{{\cal X}} 
\def\calY{{\cal Y}} 
\def\calW{{\cal W}} 
\def\calZ{{\cal Z}}

\def\tT{{\tilde T}}
\def\talpha{{\tilde\alpha}}
\def\tbeta{{\tilde\beta}}
\def\tchi{{\tilde\chi}}
\def\tdelta{{\tilde\delta}}
\def\tDelta{{\tilde\Delta}}
\def\teta{{\tilde\eta}} 
\def\tlamb{{\tilde\lambda}}
\def\tmu{{\tilde\mu}}
\def\tphibf{{\tilde\phibf}}
\def\trho{{\tilde\rho}}
\def\tvarphibf{{\tilde\varphibf}} 
\def\tq{\tilde q}
\def\tw{{\tilde\omega}}
\def\twn{{\tilde\omega_n}}
\def\twnu{{\tilde\omega_\nu}}

\def\asinh{{\rm asinh}} 
\def\Tbkt{T_{\rm BKT}}

\title{From Bose glass to many-body localization in a one-dimensional disordered Bose gas}
	
\author{Vincent Grison}	
\author{Nicolas Dupuis}
\affiliation{Sorbonne Universit\'e, CNRS, Laboratoire de Physique Th\'eorique de la Mati\`ere Condens\'ee, LPTMC, F-75005 Paris, France}
	
\date{April 2, 2026} 
	
\begin{abstract} 
We determine the finite-temperature phase diagram of a one-dimensional disordered Bose gas using bosonization and the nonperturbative functional renormalization group (RG). We discuss two different scenarios, based on distinct truncations of the effective action. In the first scenario, the Bose glass is destabilized at any finite temperature, giving rise to a normal fluid. Nevertheless, one can distinguish a low-temperature glassy regime, where disorder plays an important role on intermediate length and time scales, from a high-temperature regime, where disorder becomes irrelevant. In the second scenario, below a temperature $T_c$, the RG flow exhibits a singularity at a finite value of the RG momentum scale. We propose that this singularity signals a lack of thermalization and the existence of a localized phase for $T<T_c$. We provide a description of this low-temperature localized phase within a droplet picture and highlight a number of possible similarities with characteristics of many-body localized phases, including non-thermal behavior, avalanche instabilities and many-body resonances, the structure of the many-body spectrum, and slow dynamics in the ergodic phase. The normal fluid above $T_c$, and below a crossover temperature $T_g$, exhibits glassy properties on intermediate scales. 
\end{abstract}
	
\maketitle
	

\section{Introduction} 

Understanding the phase diagram of quantum many-body systems as a function of interactions, disorder strength, and temperature is a central question in condensed matter and in the study of ultracold atomic gases. In a one-dimensional Bose gas, disorder can destabilize the superfluid ground state (Luttinger liquid) in favor of a localized phase referred to as the Bose glass. The perturbative renormalization group (RG) predicts the superfluid--Bose-glass transition to belong to the Berezinskii-Kosterlitz-Thouless (BKT) universality class in the weak-disorder limit~\cite{Giamarchi87,Giamarchi88}. The superfluid phase is unstable in the presence of an arbitrarily weak disorder when the Luttinger parameter $K$ ---which encodes the strength of boson-boson interactions and determines the power-law decay of correlation functions--- is smaller than $3/2$. The Bose-glass phase, which also exists in higher dimensions, is characterized by a nonzero compressibility and the absence of a gap in the conductivity~\cite{Fisher89}. At finite temperature, the perturbative RG indicates that the random potential is suppressed by thermal fluctuations in the low-energy limit, and the system therefore behaves as a normal fluid~\cite{Glatz02,Glatz04}. 

Michal, Aleiner, Altshuler, and Shlyapnikov (MAAS) have proposed a markedly different scenario~\cite{Michal16}. They argue that a one-dimensional disordered Bose gas undergoes a finite-temperature fluid-insulator transition whenever the ground state is localized (i.e., is a Bose glass). This conclusion closely parallels the original many-body localization (MBL) scenario, in which an interacting quantum system at finite temperature, in the absence of coupling to an external bath, fails to thermalize and remains insulating below a critical temperature $T_c$~\cite{Basko06,Gornyi05}.  

In this paper, we determine the finite-temperature phase diagram of a one-dimensional disordered Bose gas using bosonization, the replica formalism, and the nonperturbative functional renormalization group (FRG). This approach has previously been used to study the zero-temperature phase diagram~\cite{Dupuis19,Dupuis20,Dupuis20a,Daviet20,Daviet21,Dupuis24a,Dupuis24b}. It was found that the Bose-glass phase is described by a nonperturbative fixed point characterized by a vanishing Luttinger parameter and a cuspy renormalized (functional) disorder correlator. As in classical disordered systems, in which the long-distance physics is controlled by a zero-temperature fixed point, the cuspy renormalized disorder correlator signals the presence of metastable states and glassy behavior~\cite{Dupuis19}. Moreover, the low-energy physics can be understood within the droplet picture originally proposed for spin glasses~\cite{Fisher88b}, in which low-lying excitations (droplets) couple to the (classical) ground state {\it via} quantum tunneling~\cite{Dupuis20}.  

\begin{figure}
	\centerline{\includegraphics[width=8cm]{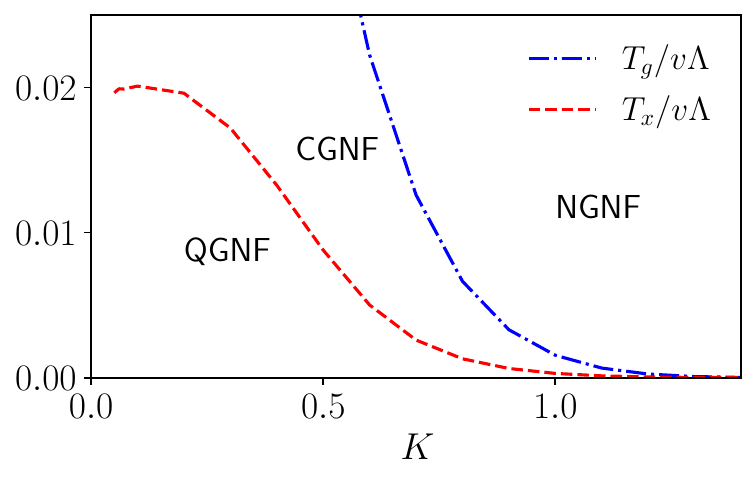}}
	\caption{Phase diagram of a one-dimensional disordered Bose gas, obtained from a truncation of the effective action that neglects the (quantum) time-derivative term in the two-replica component. The Luttinger parameter is varied at fixed dimensionless disorder (see Sec.~\ref{sec_finiteT}). The ground state is a Bose glass for $K<3/2$ (in the limit of infinitesimal disorder), while the three finite-temperature regimes correspond to the quantum glassy normal fluid (QGNF), the classical glassy normal fluid (CGNF), and the nonglassy normal fluid (NGNF). In the glassy normal-fluid regimes, the RG flow is controlled by the zero-temperature Bose-glass fixed point on intermediate length and time scales. The correlation length is set by the zero-temperature localization length $\xiloc$ in the quantum glassy regime, and by the thermal length in the classical glassy regime.
		}
	\label{fig_TgTx_W30}
\end{figure}

The FRG approach used in the following differs from previous work in two important aspects. First, we consider more general ans\"atze for the effective action (the main quantity of interest in the FRG), which include second-order space- and time-derivative terms in the two-replica component, while terms involving more than two replicas are neglected. Second, we discuss the system at finite temperature. 

The first ansatz we consider includes all second-order derivative terms in the one- and two-replica components of the effective action, except the (quantum) time-derivative term in the two-replica sector. In that case, consistent with the perturbative RG approach~\cite{Glatz02,Glatz04}, one finds that the Bose glass is destabilized at any nonzero temperature, giving rise to a normal fluid. However, three different finite-temperature regimes can be identified, as shown in Fig.~\ref{fig_TgTx_W30}. Below a crossover temperature $T_x\sim v/\xiloc$ ($v$ is the sound-mode velocity in the clean system and $\xiloc$ is the zero-temperature localization length), the FRG flow is controlled by the zero-temperature Bose-glass fixed point on intermediate length and time scales, giving rise to a quantum glassy normal fluid. The correlation length associated with density fluctuations is set by the zero-temperature localization length $\xiloc$, and we expect glassy properties on intermediate scales. The temperature range $T_x\lesssim T\lesssim T_g$ is characterized by a classical glassy regime, where the FRG flow is still controlled by the zero-temperature localized fixed point on intermediate scales, but the correlation length is determined by thermal fluctuations. The third regime, where the temperature exceeds the crossover temperature $T_g$, corresponds to a normal fluid in which disorder plays no significant role at any scale. 

\begin{figure}
\includegraphics[width=8cm]{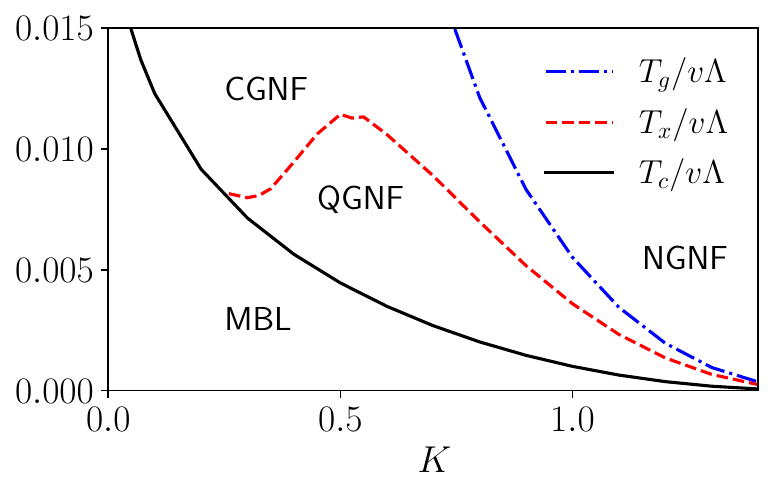}
\caption{Phase diagram obtained from a truncation of the effective action that includes all second-order derivative terms in the one- and two-replica components. The existence of a finite-temperature transition between a low-temperature localized (MBL) phase and a high-temperature normal fluid is consistent with the conclusions of Michal, Aleiner, Altshuler, and Shlyapnikov (MAAS)~\cite{Michal16}. Possible similarities between the low-temperature localized phase and the characteristic properties of MBL phases are discussed in Sec.~\ref{sec_mbl}.
}
\label{fig_TcEstim}
\end{figure}

The second ansatz includes all second-order derivative terms. It differs from the first ansatz by the inclusion of a time-derivative term in the two-replica component. In this case, already at zero temperature, we find significant differences compared to the results obtained with the first ansatz. The flow exhibits a finite-scale singularity (assuming the Luttinger parameter $K$ is smaller than $3/2$), i.e., for a nonzero value $k_c$ of the RG momentum scale $k$, with cusps appearing in the renormalized disorder correlator. We propose that this singularity is not an artifact of the truncation of the effective action, but signals a lack of thermalization at low temperatures, implying the breakdown of the (equilibrium) Matsubara formalism used in the FRG approach. We argue that this scenario can be understood within a modified droplet picture, in which quantum-active droplets ---those that couple to the ground state--- have a maximum size of order $1/k_c$. The singularity persists up to a temperature $T_c \sim v/\xiloc$ (of order the crossover temperature $T_x$ found with the first ansatz), which corresponds to a transition between a high-temperature normal-fluid phase and a low-temperature localized phase. This finite-temperature fluid-insulator transition is consistent with the results of MAAS~\cite{Michal16}. Above the transition temperature $T_c$, the finite-scale singularity disappears and the FRG flow can be integrated up to infinite length scales (i.e., $k=0$). Disorder is then irrelevant in the low-energy limit and the system behaves as a normal fluid. However, on intermediate scales, the flow is controlled by the zero-temperature localized fixed point, giving rise to the three regimes described above: quantum glassy regime ($T_c<T\lesssim T_x$), classical glassy regime ($T_x\lesssim T\lesssim T_g$), and nonglassy normal fluid ($T\gtrsim T_g$), as shown in Fig.~\ref{fig_TcEstim}. 

The outline of the paper is as follows. The FRG formalism is presented in Sec.~\ref{sec_frg}. The zero-temperature flow equations are analyzed in Sec.~\ref{sec_zeroT}, discussing both truncation schemes of the effective action introduced above. The finite-temperature flow and the phase diagram are discussed in Sec.~\ref{sec_finiteT}. Finally, we highlight a number of possible similarities between the low-temperature localized phase found with the second ansatz and certain characteristics of MBL phases: lack of thermalization, the avalanche mechanism, many-body resonances, structure of the many-body spectrum, and slow dynamics in the ergodic phase (Sec.~\ref{sec_mbl}).

\section{FRG formalism}
\label{sec_frg}

\subsection{Model} 

We consider a one-dimensional disordered Bose gas described by the Hamiltonian $\hat H_0+\Hint$ (with $\hbar=k_B=1$ throughout). Using standard bosonization techniques~\cite{Giamarchi_book}, we obtain the Hamiltonian in the absence of disorder, 
\begin{equation} 
\hat H_0 = \int dx \frac{v}{2\pi} \left[ \frac{1}{K} (\dx\hat\varphi)^2 + K (\dx\hat\theta)^2\right] ,
\label{ham}
\end{equation} 
where $\hat\theta$ is the phase of the boson operator $\hat\psi(x)=e^{i\hat\theta(x)} \hat\rho(x)^{1/2}$. The operator $\hat\varphi(x)$ is related to the density operator by 
\begin{equation} 
\hat\rho(x) = \rho_0 - \frac{1}{\pi} \dx \hat\varphi(x) + 2\rho_2 \cos[2\pi\rho_0x - 2 \hat\varphi(x)] , 
\end{equation} 
where $\rho_0$ is the average density, $\rho_2$ is a nonuniversal parameter that depends on microscopic details, and higher-order harmonics are neglected. The operators $\hat\varphi$ and $\hat\theta$ satisfy the commutation relations $[\hat\theta(x),\partial_y \hat\varphi(y)]=i\pi\delta(x-y)$. The Luttinger parameter $K$ quantifies the strength of boson-boson interactions, while $v$ denotes the sound-mode velocity. The ground state of $\hat H_0$ is a Luttinger liquid, i.e., a superfluid state with superfluid stiffness $\rho_s=vK/\pi$ and compressibility $\kappa=K/\pi v$~\cite{Giamarchi_book}.

A random potential that couples to the particle density contributes to the Hamiltonian through a term~\cite{Giamarchi87,Giamarchi88} 
\begin{equation} 
\Hdis = \int dx \biggl\{ - \frac{1}{\pi} \eta \dx \hat\varphi + \rho_2 [ \xi^* e^{i2\hat\varphi} + \hc ] \biggr\}  , 
\end{equation}  
where $\eta(x)$ (real) and $\xi(x)$ (complex) denote random potentials with Fourier components near $0$ and $\pm 2\pi\rho_0$, respectively. We assume zero-mean Gaussian distributions with the following variances,
\begin{equation}
\begin{split}
\overline{\eta(x)\eta(x')} &= D_f \delta(x-x') , \\ 
\overline{\xi^*(x)\xi(x')} &= D_b \delta(x-x') .
\end{split} 
\label{Gaussdist}
\end{equation} 
All other cumulants, e.g. $\overline{\xi(x)\xi(x')}$, vanish. An overline denotes an average over the random potentials $\eta$ and $\xi$. 

In the functional integral formalism~\cite{NDbook1}, after integrating out the field $\theta$, one obtains the Euclidean (imaginary-time) action 
\begin{align} 
S[\varphi;\eta,\xi] ={}& \int_{x,\tau} \biggl\{ \frac{v}{2\pi K} \biggl[ (\dx\varphi)^2 +  \frac{(\dtau\varphi)^2}{v^{2}} \biggr] \nonumber\\ & -\frac{1}{\pi}\eta \dx\varphi + \rho_2[ \xi^* e^{2i\varphi} + \cc ] \biggr\}  , 
\end{align}
where we use the notation $\int_{x,\tau}=\inttau\int dx$, with $\beta=1/T$, and $\varphi(x,\tau)$ is a bosonic field. The model is regularized by an ultraviolet cutoff $\Lambda$ that acts on both momenta and frequencies. The partition function 
\begin{equation}
\calZ[J;\eta,\xi] \equiv e^{W[J;\eta,\xi]} = \int \calD[\varphi] \, e^{-S[\varphi;\eta,\xi] + \int_{x,\tau} J \varphi} 
\end{equation} 
is a functional of both the external source $J$ and the random potentials $\eta$ and $\xi$. The thermodynamics of the system depends on the disorder-averaged free energy
\begin{equation} 
W_1[J] = \overline{W[J;\eta,\xi]} ,
\end{equation} 
but full information about the system requires access to higher-order cumulants of the random functional $W[J;\eta,\xi]$. In particular, the second cumulant, 
\begin{equation} 
W_2[J_a,J_b] = \overline{ W[J_a;\eta,\xi]  W[J_b;\eta,\xi] } - W_1[J_a] W_1[J_b] ,
\label{W2} 
\end{equation} 
which can be interpreted as a renormalized disorder correlator, contains information about the nature of the low-lying states of the system. 

The cumulants $W_i[J_{a_1},\dots,J_{a_i}]$ can be obtained by considering $n$ copies (or replicas) of the system~\cite{Dupuis20}. Thus, we consider 
\begin{align}
\calZ[\{J_a\}] &= \overline{ \prod_{a=1}^n \calZ[J_a;\eta,\xi]} \nonumber\\ 
&= \int \calD[\{\varphi_a\}] \, e^{-S[\{\varphi_a\}] + \sum_a \int_{x,\tau} J_a\varphi_a} ,
\label{Zrep}
\end{align} 
where the replicated action 
\begin{align}
S[\{\varphi_a\}] ={}& \sum_a \int_{x,\tau} \frac{v}{2\pi K} \left[ (\dx\varphi_a)^2 + \frac{(\dtau\varphi_a)^2}{v^2} \right] \nonumber\\ &
- \calD \sum_{a,b} \int_{x,\tau,\tau'} \cos[2\varphi_a(x,\tau)-2\varphi_b(x,\tau')] \nonumber\\ &
- \calF \sum_{a,b} \int_{x,\tau,\tau'} [\dx\varphi_a(x,\tau)][\dx\varphi_b(x,\tau')] 
\label{Srep} 
\end{align}
is obtained using~(\ref{Gaussdist}). Here, $\calF=D_f/2\pi^2$, $\calD=\rho_2^2 D_b$ and $a,b=1\dots n$ are replica indices. The functional 
\begin{align}
W[\{J_a\}] &= \ln \calZ[\{J_a\}] \nonumber\\ 
&= \sum_a W_1[J_a] + \half \sum_{a,b} W_2[J_a,J_b] + \dots 
\end{align}
is simply related to the cumulants $W_i$ of $W[J;\eta,\xi]$. In what follows, we will use the FRG to calculate the first two cumulants, $W_1$ and $W_2$. Note that, unlike the standard use of the replica trick, in which the analytic continuation $n\to 0$ opens the possibility of a spontaneous breaking of replica symmetry (a signature of glassy behavior)~\cite{Mezard_book}, in the FRG approach one usually breaks the replica symmetry explicitly by introducing $n$ external sources $\{J_a\}$ acting on each replica independently~\cite{Tarjus08}.

\subsection{Scale-dependent effective action} 

The strategy of the nonperturbative FRG approach is to build a family of models indexed by a momentum scale $k$, such that fluctuations are smoothly taken into account as $k$ is lowered from the UV scale $\Lambda$ down to zero~\cite{Berges02,Delamotte12,Dupuis_review}. This is achieved by adding to the action~(\ref{Srep}) an infrared regulator term 
\begin{equation} 
\Delta S_k[\{\varphi_a\}] = \half \sum_{a,q,\wn} |\varphi_a(q,i\wn)|^2 R_k(q,i\wn) , 
\end{equation} 
where $\wn=2\pi n/\beta$ ($n\in\mathbb{Z}$) is a Matsubara frequency. The cutoff function $R_k(q,i\wn)$ is chosen so that fluctuation modes satisfying $|q|,|\wn|/v_k\ll k$ are suppressed, while those satisfying $|q|\gg k$ or $|\wn|/v_k\gg k$ are left unaffected (the $k$-dependent sound-mode velocity $v_k$ is defined below). The precise form of the regulator will be specified later. 

The partition function therefore becomes $k$ dependent,
\begin{equation} 
\calZ_k[\{J_a\}] = \int \calD[{\varphi_a}]\, e^{-S[\{\varphi_a\}] - \Delta S_k[\{\varphi_a\}] + \sum_a \int_{x,\tau} J_a\varphi_a } .
\end{equation} 
The expectation value of the field is given by 
\begin{equation} 
\phi_{a,k}[x,\tau;\{J_f\}] = \frac{\delta \ln \calZ_k[\{J_f\}]}{\delta J_a(x,\tau)} = \mean{\varphi_a(x,\tau)} . 
\end{equation} 
The scale-dependent effective action (or Gibbs free energy),
\begin{equation} 
\Gamma_k[\{\phi_a\}] = - \ln \calZ_k[\{J_a\}] + \sum_a \int_{x,\tau} J_a\phi_a - \Delta S_k[\{\phi_a\}] ,
\label{modifiedGamma}
\end{equation} 
is defined as a modified Legendre transform, which includes the subtraction of the regulator term $\Delta S_k[\{\phi_a\}]$. Assuming that for $k=\Lambda$ the fluctuations are completely frozen by the term $\Delta S_\Lambda$, we find $\Gamma_\Lambda[\{\phi_a\}]=S[\{\phi_a\}]$, as in mean-field theory. On the other hand, the effective action of the original model~(\ref{Srep}) is recovered as $\Gamma_{k=0}$, provided that $R_{k=0}$ vanishes. The nonperturbative FRG approach aims to determine $\Gamma_{k=0}$ from $\Gamma_\Lambda$ using the Wetterich equation~\cite{Wetterich93,Ellwanger94,Morris94},
\begin{equation} 
\dt \Gamma_k[\{\phi_a\}] = \half \Tr \Bigl\{ \dt R_k \bigl( \Gamma^{(2)}_k[\{\phi_a\}] + R_k \bigr)^{-1} \Bigr\} ,
\label{eqWet} 
\end{equation} 
where $\Gamma^{(2)}_k$ is the second functional derivative of $\Gamma_k$ and $t=\ln(k/\Lambda)$ is a (negative) RG ``time''. The trace in~(\ref{eqWet}) involves a sum over momenta and frequencies, as well as over replica indices. 

The effective action 
\begin{equation} 
\Gamma_k[\{\phi_a\}] = \sum_a \Gamma_{1,k}[\phi_a] - \half \sum_{a,b} \Gamma_{2,k}[\phi_a,\phi_b] + \dots 
\label{Gammaexpansion} 
\end{equation} 
can be expanded in an increasing number of free replica sums~\cite{Tarjus08}. The minus sign in~(\ref{Gammaexpansion}) is chosen for convenience. Since $\Gamma_k[\{\phi_a\}]$ is the Legendre transform of $W_k[\{J_a\}]$, the $\Gamma_{k,i}$'s can be related to the cumulants $W_{k,i}$. The functional $\Gamma_{1,k}[\phi_a]$ is the Legendre transform of $W_1[J_a]$ and determines the thermodynamics of the system. $\Gamma_{2,k}[\phi_a,\phi_b]$ is directly identified with $W_2[J_a,J_b]$ and therefore can be considered the renormalized (functional) disorder correlator~\cite{Tarjus08,Dupuis20}.

To approximately solve the exact flow equation~(\ref{eqWet}), we retain only $\Gamma_{1,k}$ and $\Gamma_{2,k}$ in the free replica sum expansion~(\ref{Gammaexpansion}). We consider the most general ansatz up to second order in derivatives, i.e.,  
\begin{equation} 
\Gamma_{1,k}[\phi_a] = \int_{x,\tau} \frac{v_k}{2\pi K_k} \biggl[ (\dx\phi_a)^2 + \frac{(\dtau\phi_a)^2}{v_k^2} \biggr] , 
\label{ansatz1} 
\end{equation} 
and 
\begin{align}
\Gamma_{2,k}[\phi_a,\phi_b] ={}& \int_{x,\tau,\tau'} \Bigl\{ W_{1,k}(\phi_a-\phi_b) \dx\phi_a \dx\phi_b \nonumber\\& 
+ \half W_{2,k}(\phi_a-\phi_b) \bigl[ (\dx\phi_a)^2 + (\dx\phi_b)^2 \bigr] \nonumber\\& 
+ \half W_{3,k}(\phi_a-\phi_b) \bigl[ (\dtau\phi_a)^2 + (\partial_{\tau'}\phi_b)^2 \bigr] \nonumber\\& 
+ V_k(\phi_a-\phi_b) 
\Bigr\} , 
\label{ansatz2} 
\end{align} 
where $\phi_a\equiv\phi_a(x,\tau)$ and $\phi_b\equiv\phi_b(x,\tau')$. The ansatz~(\ref{ansatz1},\ref{ansatz2}) is strongly constrained by the statistical tilt symmetry (STS), 
\begin{align}
\Gamma_k[\{\phi_a\}] ={}& \Gamma_k[\{\phi_a+w\}] - \frac{n}{2} \beta (Z_x - 2n\beta\calF) \int_x (\dx w)^2 \nonumber\\ & 
- (Z_x - 2n\beta\calF) \int_{x,\tau} \sum_a (\dx w)(\dx\phi_a) , 
\end{align}  
where $w(x)$ is an arbitrary time-independent function, and $Z_x=v/\pi K$ (see Appendix~\ref{app:STS}). The STS implies 
\begin{equation} 
\frac{v_k}{\pi K_k} = \frac{v}{\pi K} = Z_x ,
\label{Zx} 
\end{equation} 
and no terms in $\Gamma_{1,k}$ other than those in~(\ref{ansatz1}) are allowed at second order in derivatives. Furthermore,
\begin{equation} 
W_{1,k}(u) + W_{2,k}(u) = 2\calF .
\label{STS}  
\end{equation} 
A term like $W_{4,k}(\phi_a-\phi_b)\dtau\phi_a\partial_{\tau'}\phi_b$ corresponds to a total derivative and can be omitted~\cite{no1}. Since $\phi_a$ is odd under parity~\cite{no2}, the two-replica effective potential $V_k$ and the functions $W_{i,k}$ ($i=1,2,3$) can only be even functions of $\phi_a-\phi_b$. They cannot depend separately on $\phi_a$ and $\phi_b$ due to the invariance of $\Gamma_{k}[\{\phi_f\}]$ under $\phi_f\to\phi_f+\const$.
 
The parameters $K_k$ and $v_k$ appearing in~(\ref{ansatz1}) can be seen as the scale-dependent Luttinger parameter and sound-mode velocity, respectively. They allow us to define a renormalized superfluid stiffness, and renormalized compressibility,
\begin{equation} 
\rho_{s,k} = \frac{v_k K_k}{\pi} , \qquad 
\kappa_k = \frac{K_k}{\pi v_k} . 
\end{equation}  
The STS implies that the compressibility $\kappa_k=K/\pi v$ is not renormalized. The two-replica effective potential $V_k(\phi_a-\phi_b)$ has been considered in previous works on the disordered one-dimensional Bose gas~\cite{Dupuis19,Dupuis20,Dupuis20a,Daviet20,Daviet21,Dupuis24a,Dupuis24b}, but the derivative terms depending on $W_{1,k}$, $W_{2,k}$, and $W_{3,k}$ have not yet been taken into account. 

The initial value $\Gamma_\Lambda$ of the effective action is defined by 
\begin{equation} 
\begin{gathered}
V_\Lambda(u) = 2\calD \cos(2u) , \quad W_{1,\Lambda}(u) = 2\calF , \\ 
W_{2,\Lambda}(u)=W_{3,\Lambda}(u)=0 , 
\end{gathered} 
\end{equation} 
and $K_\Lambda=K$, $v_\Lambda=v$. All calculations are performed with the cutoff function 
\begin{equation}
R_k(q,i\wn) = Z_x \left (q^2 + \frac{\wn^2}{v_k^2} \right) r \left( \frac{q^2+\wn^2/v_k^2}{k^2} \right), 
\label{regdef}
\end{equation}
where $r(x)=\alpha/(e^x-1)$, with $\alpha$ a free parameter of order unity (in practice we take $\alpha=2$). 

It should be noted that the case of hard-core bosons, where $K=1$ and the Hamiltonian $\hat H_0+\Hdis$ also describes a system of noninteracting fermions in a random potential, is a special case due to the absence of inelastic processes. In that case, care must be taken not to introduce inelastic processes into the renormalization procedure~\cite{Giamarchi88}. The cutoff function~(\ref{regdef}), which acts on both momenta and frequencies, is not suitable for this case. In the following, we do not consider hard-core bosons. The case $K=1$ (and more generally $K\leq 1$) can nevertheless be realized, but corresponds to a system of interacting bosons with finite-range interactions. 

\subsection{Dimensionless variables} 

It is convenient to write the effective action in dimensionless form, using the variables
\begin{equation} 
\tilde x = kx, \quad \tilde\tau = v_k k \tau . 
\end{equation} 
If we assign the scaling dimension $[x]=-1$ to the space coordinate, the time coordinate has scaling dimension $[\tau]=-z_k$, where the (running) dynamical critical exponent $z_k$ is defined by $\dt v_k=(z_k-1)v_k$ with $\dt=k\dk$. In terms of the dimensionless variables $\tilde x,\tilde\tau$, the one-replica effective action reads 
\begin{equation} 
\Gamma_{1,k}[\phi_a] = \frac{1}{2\pi K_k} \int_{\tilde x,\tilde\tau} \bigl[ (\dxtilde\phi_a)^2 + (\dtautilde\phi_a)^2 \bigr] ,
\label{ansatz1a} 
\end{equation} 
where $\tilde\tau\in [0,\tilde\beta_k]$, with $\tilde\beta_k=1/\tilde T_k$, and 
\begin{equation}
\tilde T_k = \frac{T}{v_kk} 
\label{tildeTk}
\end{equation} 
is the dimensionless temperature. Note that we do not rescale the field $\phi_a$, which is dimensionless. $K_k$ plays the role of the ``coupling constant'', whereas $v_k$ has been eliminated {\it via} a suitable definition of the dimensionless time $\tilde\tau$.  In addition to the dynamical critical exponent $z_k$, we define the exponent $\theta_k$ by $\dt K_k=\theta_k K_k$. Equation~(\ref{Zx}) implies that $\theta_k$ and $z_k$ are not independent: $\theta_k=z_k-1$. 

The effective action $\Gamma_{1,k}$ has a scaling dimension of $1-z_k$. It is natural to require the second cumulant $\Gamma_{2,k}$ to have a scaling dimension $[\Gamma_{2,k}]=2[\Gamma_{1,k}]=2(1-z_k)$. This implies $[V_k]=3$, $[W_{1,k}]=[W_{2,k}]=1$, and  $[W_{3,k}]=3-2z_k$. Introducing the dimensionless functions
\begin{equation} 
\begin{split}
\tilde V_k(u) &= \frac{K^2}{v^2 k^3} V_k(u) , \\ 
\tilde W_{i,k}(u) &= \frac{K^2}{v^2 k} W_{i,k}(u) \quad (i=1,2) , \\  
\tilde W_{3,k}(u) &= \frac{K^2 v_k^2}{v^2 k} W_{3,k}(u) , 
\end{split} 
\label{dimpotdef} 
\end{equation} 
we write the two-replica part of the effective action as 
\begin{align}
\Gamma_{2,k}[\phi_a,\phi_b] ={}& \frac{1}{K_k^2} \int_{\tilde x,\tilde\tau,\tilde\tau'}  \Bigl\{ \tilde W_{1,k}(\phi_a-\phi_b) \partial_{\tilde x} \phi_a \partial_{\tilde x}\phi_b \nonumber\\& 
+ \half \tilde W_{2,k}(\phi_a-\phi_b) \bigl[ (\partial_{\tilde x}\phi_a)^2 + (\partial_{\tilde x}\phi_b)^2 \bigr] \nonumber\\& 
+ \half \tilde W_{3,k}(\phi_a-\phi_b) \bigl[ (\partial_{\tilde\tau}\phi_a)^2 + (\partial_{\tilde\tau'}\phi_b)^2 \bigr] \nonumber\\& 
+ \tilde V_k(\phi_a-\phi_b) 
\Bigr\} .
\label{ansatz2a} 
\end{align}
Instead of $\tilde V_k$, in the following we consider its second derivative $\tilde\Delta_k(u)=-\tilde V_k''(u)$.

\subsection{Flow equations} 

The flow equations for $\theta_k=z_k-1$ and the functions $\tilde V_k$, $\tilde W_{1,k}$, $\tilde W_{2,k}$, and $\tilde W_{3,k}$, obtained by inserting the ansatz~(\ref{ansatz1},\ref{ansatz2}) into the Wetterich equation~(\ref{eqWet}), are given in Appendix~\ref{app:floweq}. They can be written in the form
\begin{align}
    \theta_k ={}& \frac{\pi}{2} \, \bigl[ \bar M^\tau_{0,1}(0)   \tilde\Delta_k''(0)   +   \bar M^\tau_{0,1}(1)   \tilde{W}_{1,k}''(0)  \nonumber\\ & -   \bar l_1(0)   \tilde{W}_{3,k}''(0)   -   8   \tilde\Omega^2   \bar M^\tau_{0,1}(0)   \tilde{W}_{3,k}''(0) \bigr]  
\end{align}
(with $\tilde\Omega=2\pi \tilde T_k$) and 
\begin{align}
\dt \tilde \Delta_k(u) ={}& -3 \tilde\Delta_k(u) - K_k l_1(0,0) \tilde\Delta''(u) \nonumber\\ & -  K_k l_1(1,0) \tilde W_{1,k}''(u) \nonumber\\ & +  K_k l_1(0,1) \tilde W_{3,k}''(u)  + \calR_k , \label{eq1} \\ 
\dt \tilde W_{1,k}(u) ={}& - \tilde W_{1,k}(u) - K_k l_1(0,0) \tilde W_{1,k}''(u) + \calR_{1,k} , \nonumber\\  
\dt \tilde W_{3,k}(u) ={}& (2z_k-3) \tilde W_{3,k}(u) \nonumber\\ & - K_k l_1(0,0) \tilde W_{3,k}''(u)  + \calR_{3,k} \nonumber ,
\end{align}  
where we have eliminated $\tilde W_{2,k}$ using~(\ref{STS}). The ``threshold'' functions $\bar M^\tau_{0,1}(n)$, $\bar l_1(0)$, and $l_1(n,m)$ are defined in Appendix~\ref{app:threshold}. Except for $\bar l_1(0)$, they depend on $k$. The overline means that $\bar M^\tau_{0,1}$ and $\bar l_1(0)$ only involve an integration over momentum; at zero temperature they depend only on the static component ($\wn=0$) of the propagator. In Eqs.~(\ref{eq1}), we have written all terms that explicitly depend on $K_k$. The ``remaining'' terms $\calR_k$ and $\calR_{1,k}$ depend only on static threshold functions and are independent of $K_k$ and $\tilde T_k$, while $\calR_{3,k}$ depends on $\tilde T_k$ through finite differences of Matsubara frequencies. $\calR_k$ and $\calR_{1,k}$ involve the functions $\tilde\Delta_k$ and $\tilde W_{1,k}$ (and their derivatives) but are independent of $\tilde W_{3,k}$. Thus, $\tilde W_{3,k}$ does not enter the flow equation for $\tilde W_{1,k}$, and appears in that of $\tilde\Delta_k$ only through the term $K_k l_1(0,1) \tilde W_{3,k}''(u)$.

For the numerical solution of the flow equations, it is convenient to use a circular-harmonic expansion
\begin{equation}
\begin{split} 
\tilde\Delta_k(u) &= \sum_{n=1}^{n_{\rm max}} \tilde\Delta_{n,k} \cos(2nu) , \\ 
\tilde W_{i,k}(u) &= \sum_{n=0}^{n_{\rm max}} \tilde W_{i,n,k} \cos(2nu) \quad (i=1,2,3) , 
\end{split}
\end{equation} 
with $n_{\rm max}$ typically of the order of 100. The zeroth-order harmonic $\tilde\Delta_{0,\Lambda}$ of the potential $\Delta_\Lambda(u)$ vanishes, a property that is preserved by the flow, 
\begin{equation} 
\tilde\Delta_{0,k} = \int_0^\pi \frac{du}{\pi}\, \tilde\Delta_k(u) = 0 .
\label{potcond} 
\end{equation}

\section{Zero-temperature flow} 
\label{sec_zeroT} 

In this section, we consider the zero-temperature limit. The analysis of the stability of the superfluid phase with respect to infinitesimal disorder is not modified by the derivative terms in $\Gamma_{2,k}$. The superfluid--Bose-glass transition can be analyzed by considering only $K_k$ and the dimensionless renormalized disorder variance $\tilde\calD_k=K^2\calD_k/v^2k^3$. For vanishing disorder ($\tilde\calD_k=0$), the transition occurs at the critical value $K=3/2$ of the Luttinger parameter, and belongs to the universality class of the BKT transition~\cite{Giamarchi87,Giamarchi88,Dupuis20}. In the following, we consider only the flow in the Bose-glass phase. All results discussed in this section and the following ones are obtained with $\calF=0$. 

\subsection{Flow without $\tilde W_{3,k}$} 
\label{subsec_Tzero_withoutW3} 

\begin{figure}
\includegraphics[width=8cm]{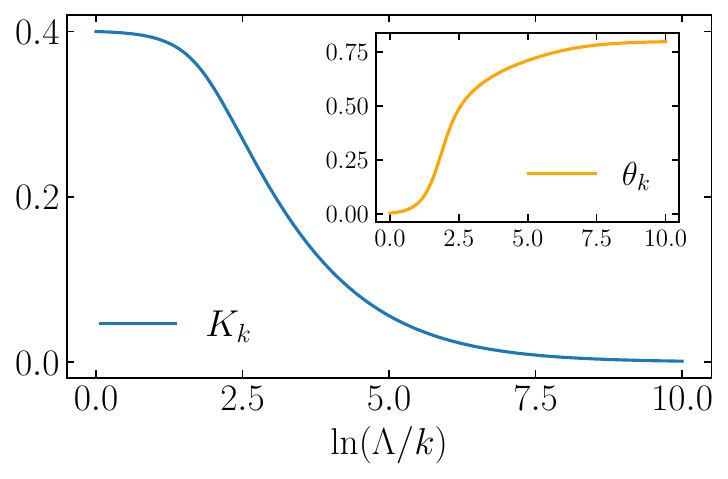}
\caption{Flow of the Luttinger parameter $K_k$ and the exponent $\theta_k=z_k-1=\dt \ln K_k$ (inset) as functions of $\ln(\Lambda/k)$ for $\tilde W_{3,k}=0$ and $T=0$ ($\tilde \Delta_{1,\Lambda}=0.002$) . } 
\label{fig_thetaK_flowW12} 
\end{figure}

We first study the flow in the absence of $\tilde W_{3,k}$. The derivative terms $\tilde W_{1,k}$ and $\tilde W_{2,k}$ do not qualitatively change the behavior obtained when only $\tilde\Delta_k$ is considered~\cite{Dupuis20}. The Luttinger parameter $K_k\sim k^\theta$ and the velocity $v_k \sim k^\theta$ vanish in the limit $k\to 0$, with an exponent $\theta=\lim_{k\to 0}\theta_k$, and the dynamical critical exponent takes the value $z=\lim_{k\to 0}z_k=1+\theta$ (Fig.~\ref{fig_thetaK_flowW12}). This implies that the superfluid stiffness $\rho_{s,k}\sim k^{2\theta}$ vanishes as $k\to 0$.

\begin{figure}
    \hspace*{-1.0cm}
    \includegraphics[width=8cm]{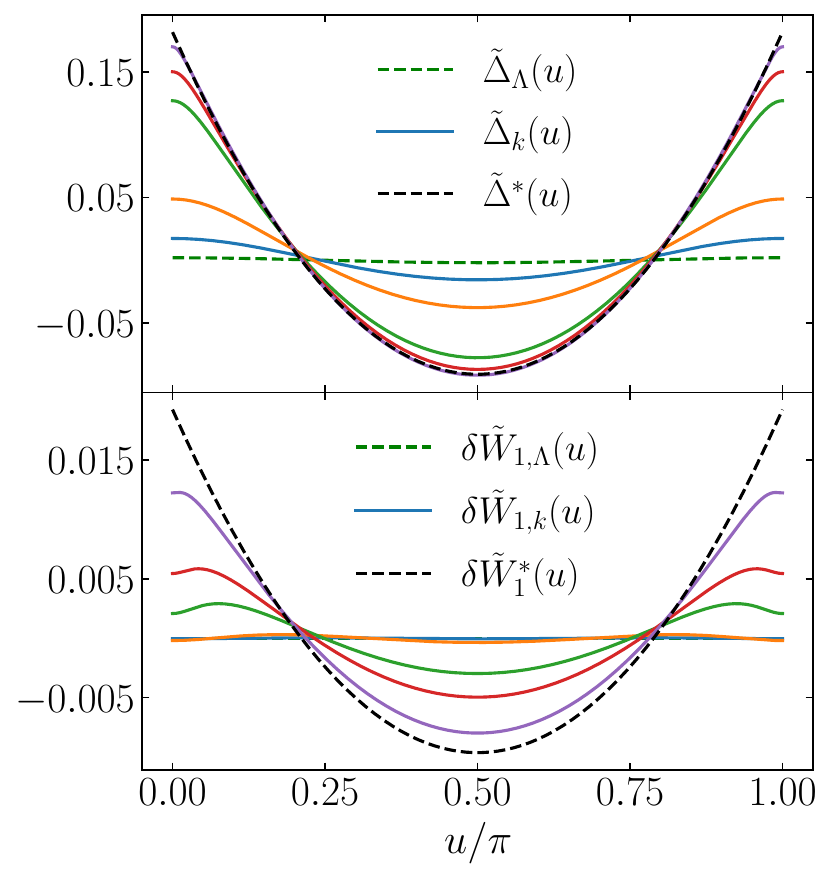}
	\caption{$\tilde\Delta_k(u)$ and $\delta\tilde W_{1,k}(u)=-\delta\tilde W_{2,k}(u)$ for $\ln(\Lambda/k)=0/1.0/1.54/2.34/2.75/3.82$ (from top to bottom for $u=\pi/2$), for $\tilde W_{3,k}=0$ and $T=0$ ($K=0.4$, $\tilde \Delta_{1,\Lambda}=0.002$). The dash-dotted (green) curves show the initial conditions $\tilde\Delta_\Lambda(u)=2\tilde\calD_\Lambda \cos(2u)$, $\delta\tilde W_{1,\Lambda}(u)=0$, and the dashed (black) curves show the fixed-point solutions~(\ref{fppot}).}
	\label{fig_pot_flowW12} 
\end{figure}

All harmonics have a well-defined limit as $k\to 0$, except the zeroth-order harmonics $\tilde W_{1,0,k}$ and $\tilde W_{2,0,k}$, which diverge and satisfy  
\begin{equation} 
\tilde W_{1,0,k} + \tilde W_{2,0,k} = 2 \tilde \calF_k = 2 \frac{K^2\calF}{v^2k}  
\end{equation} 
(in agreement with~(\ref{STS})), but their flow does not affect the other harmonics. 
This means that the functions $\tilde\Delta_k(u)$ and
\begin{equation}
\delta\tilde W_{i,k}(u)=\tilde W_{i,k}(u)-\tilde W_{i,0,k} \quad (i=1,2)
\end{equation} 
approach fixed-point values $\tilde\Delta^* (u)$ and $\delta\tilde W^*_i(u)$ as $k\to 0$. The latter are $\pi$-periodic functions given by parabolas on the interval $[0,\pi]$: 
\begin{equation} 
\begin{split} 
\tilde\Delta^*(u) &= a \left[ u(\pi-u) - \frac{\pi^2}{6} \right] , \\    
\delta\tilde W_1^*(u) &= -\delta\tilde W_2^*(u) = b \left[ u(\pi-u) - \frac{\pi^2}{6} \right] .   
\end{split}
\label{fppot} 
\end{equation}  
The function $\tilde\Delta^*(u)$ depends on a single parameter since it must satisfy~(\ref{potcond}), while $\delta\tilde W_1^*(u)$ and $\delta\tilde W_2^*(u)$ satisfy this constraint by construction. The coefficients $a$ and $b$ can be found by inserting~(\ref{fppot}) into the equation $\dt\tilde\Delta_k(u)=\dt\delta \tilde W_{1,k}(u)=0$; this yields perfect agreement with the numerical solution of the flow equation (Fig.~\ref{fig_pot_flowW12}). For nonzero but small momentum scale ($k>0$), the cusp singularities at $u=p\pi$ ($p\in\mathbb{Z}$) are rounded into a quantum boundary layer (QBL): For values of $u$ near zero, $\tilde\Delta_k(u)-\tilde\Delta_k(0)\propto -|u|$ except within a boundary layer of size $|u|\sim K_k$. The function $\delta\tilde W_{1,k}(u)-\delta\tilde W_{1,k}(0)$ exhibits a similar behavior, although the QBL size is not simply related to $K_k$ due to the change in curvature near $u=0$ (see Fig.~\ref{fig_pot_flowW12}).

The physical content of this kind of fixed point has been discussed in Refs.~\cite{Dupuis19,Dupuis20}. The vanishing of the Luttinger parameter $K_k$, which controls quantum fluctuations of the density field $\varphi$, implies that the phase field $\varphi(x,\tau)\simeq \varphi(x)$ has weak temporal (quantum) fluctuations and adjusts to minimize the energy due to the random potential, a distinctive feature of pinning (which corresponds to a localized state for the bosons). 

On the other hand, the cuspy nonanalytic form of the two-replica potential $\tilde\Delta^*(u)$ is characteristic of many classical disordered systems, such as elastic manifolds, charge-density waves, and the random-field Ising model~\cite{Narayan92,Narayan92a,Balents93,Ledoussal04,Tissier08}. These systems are controlled by a zero-temperature fixed point, and the cusp signals the existence of many metastable states, leading to ``shock'' singularities (or static avalanches)~\cite{Balents96,Tissier08,Ledoussal09}: When the system is subjected to an external force, the ground state varies discontinuously whenever it becomes degenerate with a metastable state (which then becomes the new ground state). At finite temperature, the system has a nonnegligible probability of being in two distinct (nearly degenerate) configurations, and the discontinuity is smeared on a scale set by the temperature $T$. This explains why the cusp in the disorder correlator $\Delta_k(u)$ is rounded into a thermal boundary layer at nonzero scales $k>0$, where the (renormalized) temperature $T_k$ is nonzero. In these classical systems, the metastable states are responsible for glassy properties: pinning, ``shocks'' and ``avalanches'', chaotic behavior, aging, etc.

A similar interpretation holds in the zero-temperature Bose-glass phase, where thermal fluctuations (controlled by temperature) are replaced by quantum fluctuations (controlled by the Luttinger parameter). In the semiclassical limit $K\to 0$, the ground state evolves under a slowly varying external source with abrupt switches at discrete, sample-dependent values of the source ---corresponding to level crossings where a metastable state becomes the new ground state. It has been argued that low-lying excited states arise from soliton-antisoliton (i.e., kink-antikink) pairs of the field $\varphi$ with anomalously small excitation energy~\cite{Rosenow06,Nattermann07}. These metastable states are obtained from the (classical) ground state by shifting $\varphi$ by $\pm\pi$ in a finite region of size $L$ (the distance between the soliton and the antisoliton), corresponding to the hopping of a boson (or particle-hole pair creation) across this distance~\cite{[{The ability of the FRG to capture soliton-like excitations in one-dimensional quantum systems has been demonstrated in }]Daviet19}. For a finite Luttinger parameter $K$, quantum tunneling between the ground state and these rare metastable states is due to instantons that describe the spontaneous formation of these soliton-antisoliton pairs, which are responsible for the QBL. As in classical disordered systems governed by a zero-temperature fixed point, we expect the cuspy fixed point~(\ref{fppot}) to be associated with glassy properties~\cite{Dupuis19,Dupuis20}.  

This discussion can be rephrased in a fully quantum picture. Quantum fluctuations between the classical ground state and the low-lying metastable states give rise to a (renormalized) ground state and low-energy quantum states that can be viewed as consisting of a quantum soliton and antisoliton at a distance $L$ apart. The energy of the various quantum states, and in particular the ground state, depends on the external source. If the source Hamiltonian commutes with the system Hamiltonian, then discontinuous changes in the ground state, i.e., first-order quantum phase transitions due to level crossings, will occur when the source is varied. In a disordered macroscopic system and for a generic external perturbation, the source and system Hamiltonians are not expected to commute. In that case, as the source is varied, level crossings will be avoided, thereby suppressing the cuspy behavior of the ground-state energy.

In classical disordered systems, where the long-distance physics is controlled by a zero-temperature fixed point, the low-energy (glassy) properties are usually explained within the framework of the droplet picture~\cite{Fisher88b}. The latter assumes the existence, at each length scale $L$, of a small number of excitations above the ground state, drawn from an energy distribution $P_L(E)$ of width $\Delta E\sim L^\theta$, with a weight $\sim L^{-\theta}$ near $E=0$. The number of thermally active excitations is therefore $\sim T/L^\theta$, i.e., the system has a probability $\sim T/L^\theta$ of occupying two nearly degenerate configurations. Thermal fluctuations are dominated by these rare droplet excitations. From a theoretical point of view, the  key connection between the FRG and the droplet phenomenology is the existence of a thermal boundary layer in the disorder correlator~\cite{Balents04,Balents05}.

Given the analogy with classical disordered systems discussed above, we expect droplet phenomenology to be applicable to the Bose-glass phase~\cite{Dupuis20}. The droplets are naturally identified with the metastable states that arise from creating a soliton-antisoliton pair. They are two-dimensional since the action of the field $\varphi(x,\tau)$ is defined in a $(1+1)$-dimensional spacetime. For a droplet of spatial size $L$, the extension $L_\tau\sim L^z$ in the imaginary-time direction can be interpreted as a quantum coherence time. The number of quantum-mechanically active droplets is $\sim K/L^\theta$, i.e., $\sim Kk^\theta\sim K_k$ if we identify $1/L$ with the running RG momentum scale $k$. The FRG calculation of correlation functions supports the validity of the droplet picture in cases where the low-energy behavior of the Bose glass is controlled by the fixed point obtained when $\tilde W_{3,k}$ is omitted~\cite{Dupuis20}.

\subsection{Flow with $\tilde W_{3,k}$} 
\label{subsec_Tzero_withW3} 

\begin{figure}
\includegraphics[width=8cm]{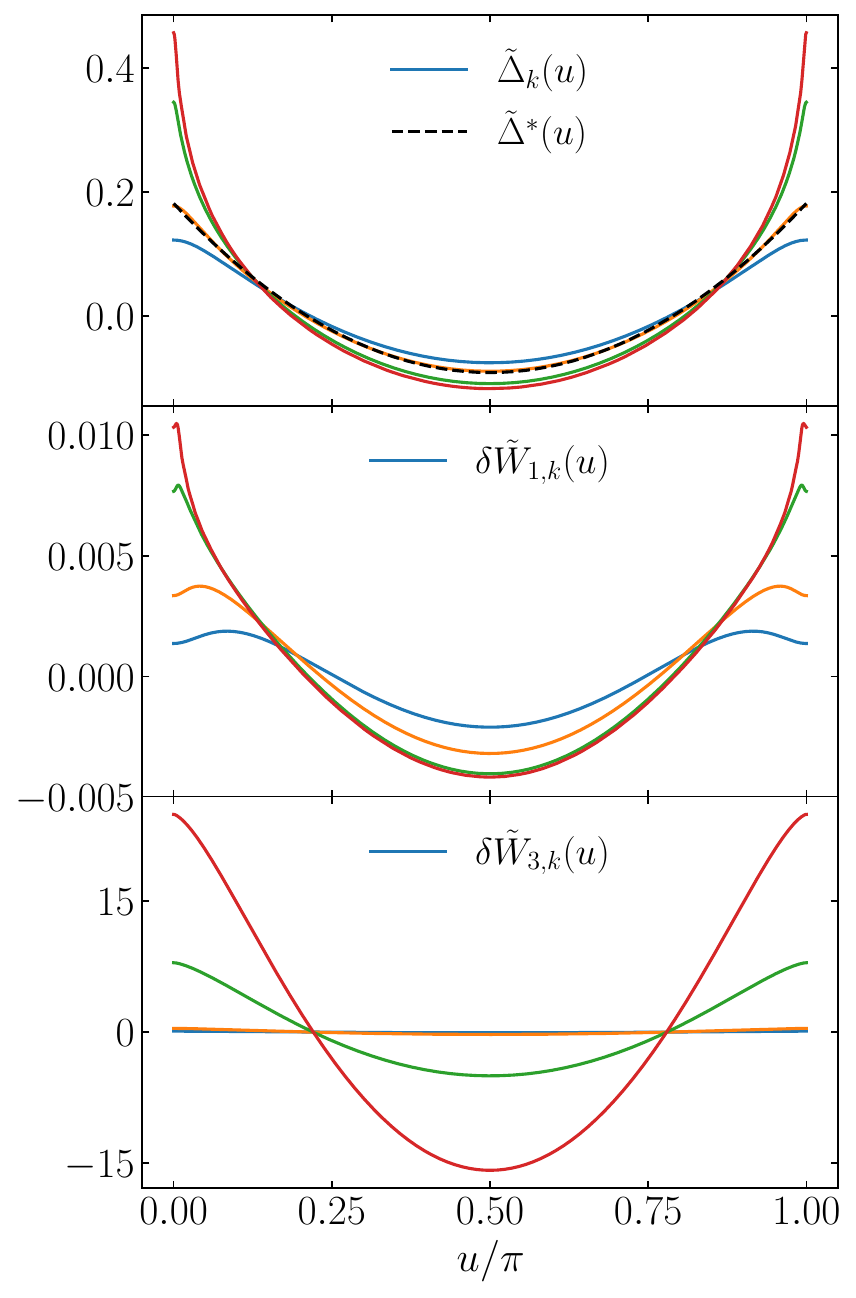}
\caption{Dimensionless disorder correlator $\tilde\Delta_k(u)$, and the derivative terms $\delta\tilde W_{1,k}(u)$ and $\delta\tilde W_{3,k}(u)$ for $\ln(\Lambda/k)\simeq 4.504/4.655/4.693/4.694$ (from top to bottom for $u=\pi/2$) at zero temperature and near the singularity at $k_c$ ($\tilde \Delta_{1,\Lambda}=0.02$, $K=1.4$). As the singularity is approached, cusps form in $\tilde\Delta_k(u)$ and $\delta\tilde W_{1,k}(u)$ around $u=0$ and $u=\pi$, and $\delta\tilde W_{3,k}(u)$ grows large. The dashed (black) line shows the fixed-point solution $\tilde\Delta^*(u)$ obtained in the absence of $\tilde W_{3,k}(u)$ [Eq.~(\ref{fppot})]. }
\label{fig_DeltaW_zerotT_kc}  
\end{figure}

\begin{figure}
	\includegraphics[width=8cm]{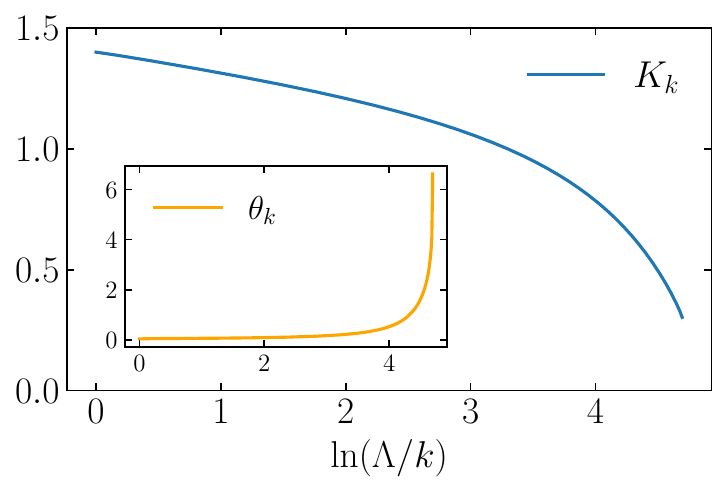}
	\caption{Flow of the Luttinger parameter $K_k$ as a function of $\ln(\Lambda/k)$ when $\tilde W_{3,k}$ is taken into account ($T=0$, $\tilde \Delta_{1,\Lambda}=0.02$, $K=1.4$). The inset shows the exponent $\theta_k=\dt\ln K_k$.}
	\label{fig_KzeroT}  
\end{figure}

The flow changes significantly when $\tilde W_{3,k}$ is taken into account. Although it is initially zero, $\tilde W_{3,k}$ is generated by the RG equations. At a finite scale, $\tilde\Delta_k(u)$ becomes nearly identical to its fixed-point value $\tilde \Delta^*(u)$ obtained for $\tilde W_{3,k}(u)=0$ (Sec.~\ref{subsec_Tzero_withoutW3}), while $\delta\tilde W_{3,k}(u)$ grows significantly. In particular, cusps form at $u=p\pi$ ($p\in\mathbb{Z}$) in $\tilde \Delta_k(u)$. These cusps sharpen quickly, and a singularity occurs at $k=k_c$, where $\delta\tilde W_{3,k}(u)$ appears to diverge (Fig.~\ref{fig_DeltaW_zerotT_kc}). The exponent $\theta_k$ and the dynamical critical exponent $z_k=\theta_k+1$ also take very large values (whether or not they actually diverge is not clear from the numerics). This corresponds to a very rapid decrease of the Luttinger parameter $K_k$, which nevertheless remains finite at $k_c$ (Fig.~\ref{fig_KzeroT}). 

It is not possible to find a numerical solution of the flow equations for $k<k_c$, regardless of the choice of numerical solver. Moreover, more elaborate approximation schemes, used to solve the exact flow equation satisfied by the effective action, and discussed in Appendix~\ref{app_beyondDE}, do not eliminate the singularity. In the following, we adopt the perspective that the finite-scale singularity at $k_c$ has a genuine physical meaning and discuss its consequences. A possible explanation for its origin, in relation to MBL, is discussed in Sec.~\ref{sec_mbl}. In any case, we do not question the localized nature of the ground state. We will see in Sec.~\ref{sec_finiteT} how the singularity manifests itself when approached from the high-temperature region. 

As mentioned in Sec.~\ref{subsec_Tzero_withoutW3}, the cusps in $\tilde\Delta_k(u)$ can be interpreted as a signature of the existence of metastable states, and the rounding into a boundary layer prior to full cusp formation is due to quantum tunneling between the ground state and metastable states. These low-lying metastable states can be seen as quantum-mechanically active droplets. In the absence of $\tilde W_{3,k}(u)$, quantum-mechanically active droplets exist on all length scales, which is reflected in the power-law vanishing of the Luttinger parameter $K_k\sim k^\theta$ as $k\to 0$. 

This interpretation must be reconsidered when $\tilde W_{3,k}(u)$ is taken into account. The cusp formation in $\tilde\Delta_k(u)$ at scale $k_c$ suggests that there are no quantum-mechanically active droplets beyond the characteristic length scale $\xi_c\sim 1/k_c$. This hypothesis is further discussed in Sec.~\ref{subsec_avalanches}. Importantly, the term of the effective action responsible for the finite-scale singularity at $k_c$ depends on the time derivative $\dtau\phi_a(x,\tau)$ of the field, and is therefore inherently quantum in nature. 

It is tempting to see the apparent divergence of $\theta_k$ at $k_c$, while $K_k$ remains finite, as a sign of a discontinuous jump of the Luttinger parameter from $K_{k_c^+}>0$ to $K_{k_c^-}=0$. If one sets $K_k$ to zero in the flow equations for $k<k_c$, one finds that $\tilde\Delta_k(u)$ and $\delta\tilde W_{1,k}(u)$ decouple from $\tilde W_{3,k}(u)$ (which then plays no role) and rapidly reach their fixed-point values~(\ref{fppot}) obtained in Sec.~\ref{subsec_Tzero_withoutW3}. Whether this fixed point is actually reached remains an open question; in the following, we will not discuss the flow beyond the singularity, i.e., for $k<k_c$, which is inaccessible within the derivative expansion~(\ref{ansatz1},\ref{ansatz2}) of the effective action (see also the discussion in Appendix~\ref{app_beyondDE}). We will see that at sufficiently high temperatures, when the singularity disappears, the fixed point~(\ref{fppot}) obtained when $\tilde W_{3,k}(u)=0$ plays an important role in the flow when $k$ is smaller than the zero-temperature singularity scale $k_c$.

\section{Finite-temperature flow and phase diagram}
\label{sec_finiteT} 

In this section, we discuss the FRG flow as a function of the initial conditions defined by $K$ and $T$. We work at fixed value of the dimensionless disorder strength, i.e., at fixed $\tilde\Delta_{1,\Lambda}$ or $\calD K^2$, where $\calD=\rho_2^2 D_b$ is the variance of the disorder. This ensures a well-defined limit as $K\to 0$, characterized by a finite value of the zero-temperature localization length. As in Sec.~\ref{sec_zeroT}, we consider only the case where the system is in the Bose-glass phase at zero temperature.

\subsection{Quantum-classical crossover} 
\label{subsec_Xover} 

At sufficiently high temperatures, the flow is regular even in the presence of $\tilde W_{3,k}$, and can be integrated from $k=\Lambda$ down to $k=0$. It always ends in a classical regime, where quantum fluctuations are suppressed. The quantum-classical crossover scale $k_x$ can be defined by the criterion $\tw_{n=1}=2\pi\tilde T_{k}=1$, i.e., 
\begin{equation} 
\tilde T_{k_x} = \frac{1}{2\pi} ,
\label{tildeTkx} 
\end{equation} 
where $\tilde T_k$ is the dimensionless temperature~(\ref{tildeTk}). 
This leads to 
\beq 
k_x = \frac{2\pi T}{v_{k_x}} = \frac{2\pi T}{v} \frac{K}{K_{k_x}} . 
\eeq 
When $k\ll k_x$, i.e. $\tilde T_k\gg 1/2\pi$, only the zero Matsubara frequency of the propagator gives a significant contribution to the threshold functions that determine the FRG flow (Appendix~\ref{app:classical_limit}). In this regime, the imaginary-time dependence of the field can therefore be neglected, so that the effective action becomes independent of $W_{3,k}$, i.e., 
\beq 
\Gamma_{1,k}[\phi_a] = \frac{Z_x}{2T} \int_x (\dx \phi_a)^2 ,
\eeq 
and
\begin{align}
\Gamma_{2,k}[\phi_a,\phi_b] ={}& \frac{1}{T^2} \int_x 
\Bigl\{ W_{1,k}(\phi_a-\phi_b) \dx\phi_a \dx\phi_b \nonumber\\& 
+ \half W_{2,k}(\phi_a-\phi_b) \bigl[ (\dx\phi_a)^2 + (\dx\phi_b)^2 \bigr] \nonumber\\& 
+ V_k(\phi_a-\phi_b) 
\Bigr\} ,
\end{align}
The one-replica component of the effective action now carries the scaling dimension $[\Gamma_{1,k}]=1$. Demanding that $[\Gamma_{2,k}]=2[\Gamma_{1,k}]=2$, we obtain $[V_k]=3$ and $[W_{1,k}]=[W_{2,k}]=1$, so that we can keep the definition~(\ref{dimpotdef}) of the dimensionless functions $\tilde V_k$, $\tilde W_{1,k}$, and $\tilde W_{2,k}$. This leads to the dimensionless form of the effective action, 
\begin{align}
\Gamma_{1,k}[\phi_a] ={}& \frac{1}{2\tilde t_k} \int_{\tilde x} (\partial_{\tilde x}\phi_a)^2 , \\ 
\Gamma_{2,k}[\phi_a,\phi_b] ={}& \frac{\pi^2}{\tilde t_k^2} \int_{\tilde x} 
\Bigl\{ \tilde W_{1,k}(\phi_a-\phi_b) \partial_{\tilde x}\phi_a \partial_{\tilde x}\phi_b \nonumber\\& 
+ \half \tilde W_{2,k}(\phi_a-\phi_b) \bigl[ (\partial_{\tilde x}\phi_a)^2 + (\partial_{\tilde x}\phi_b)^2 \bigr] \nonumber\\& 
+ \tilde V_k(\phi_a-\phi_b) 
\Bigr\} ,
\end{align} 
where 
\begin{equation} 
\tilde t_k = \frac{T}{Z_xk} =  \pi K_k \tilde T_k . 
\label{tildetk}
\end{equation} 
The coupling $\tilde t_k$, which is given by the product of the $T=0$ coupling constant $K_k$ and the dimensionless temperature $\tilde T_k$~\cite{[{A similar relation between the $T=0$ coupling constant, the renormalized temperature and the coupling constant in the classical regime is obtained in the quantum nonlinear O(3) sigma model, see e.g. }]Chakravarty89}, defines the dimensionless (renormalized) temperature in the classical regime. In the following, we will refer to both $\tilde T_k$ and $\tilde t_k$ as renormalized dimensionless temperatures, since in general no ambiguity arises.   

In the classical regime of the flow, where $\tilde T_k\gg 1/2\pi$, the temperature-dependent threshold function $l_1(a,b)$ becomes proportional to $\tilde T_k$ (Appendix~\ref{app:threshold}), 
\begin{equation} 
l_1(a,b) \simeq \delta_{b,0} \tilde T_k \bar l_1(a) , 
\end{equation} 
where $\bar l_1(a)$ is a ``classical'' (or static) threshold function that involves only momentum integration and is independent of temperature. The flow equations~(\ref{eq1}) then take the form
\begin{equation} 
\begin{split}
\dt \tilde \Delta_k(u) =& -3 \tilde\Delta_k(u) - \frac{\tilde t_k}{\pi} \bar l_1(0) \tilde\Delta''(u) + \calR_k, \\ 
\dt \tilde W_{1,k}(u) =& - \tilde W_{1,k}(u) - \frac{\tilde t_k}{\pi} \bar l_1(0) \tilde W_{1,k}''(u) + \calR_{1,k} .
\end{split}
\label{eqcl}
\end{equation}  
The contributions $\calR_k$ and $\calR_{1,k}$ are not modified by the classical limit, since they depend only on static (i.e., temperature-independent) threshold functions. Moreover, since $l_1(0,1)$ vanishes in the classical regime, $\tilde W_{3,k}(u)$ does not appear in the flow equation for $\tilde\Delta_k$. Thus, $\tilde W_{3,k}(u)$ no longer plays a role, as expected when the time dependence of the field $\phi_a(x,\tau)$ can be ignored. 

Equations~(\ref{eqcl}) admit a zero-temperature fixed point defined by $\tilde t_k=0$ and Eqs.~(\ref{fppot}), which is identical to the fixed point obtained from the zero-temperature flow equations in the absence of $\tilde W_{3,k}$ when $K_k=0$  (Sec.~\ref{subsec_Tzero_withoutW3}). However, while this fixed point is attractive at zero temperature, it is unstable at finite temperature because $\tilde t_k$ is a relevant variable. Since $\tilde t_k\sim 1/k$ eventually becomes very large, the functions $\tilde\Delta_k$ and $\tilde W_{1,k}$ become irrelevant ($\bar l_1(0)$ is positive)~\cite{no3}. Thus, the classical flow is ultimately attracted to the normal-fluid fixed point, defined by $\tilde\Delta^*(u)=\tilde W_{1}^*(u)=0$ and an infinite dimensionless temperature $\tilde t^*$. Nevertheless, we will see in the following sections that the zero-temperature fixed point~(\ref{fppot}), hereafter referred to as the localized fixed point, plays an important role at finite temperature, both without and with the $\tilde W_{3,k}$ contribution.  

\begin{figure}
\includegraphics[width=6.5cm]{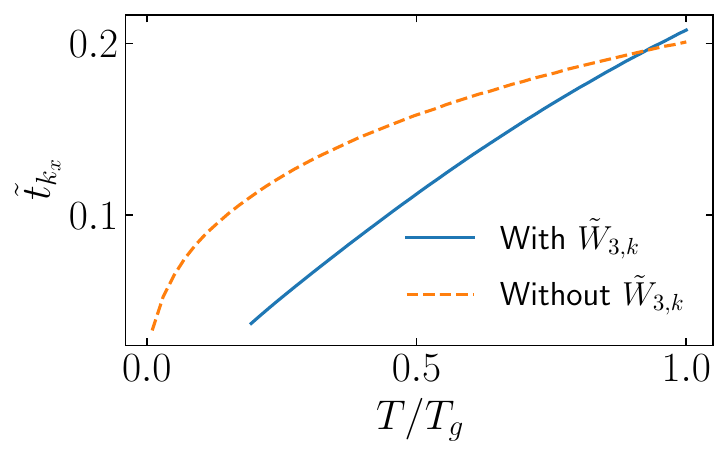}
\caption{Renormalized dimensionless temperature $\tilde t_{k_x}$ as a function of $T/T_g$ ($K=1.4$, $\tilde \Delta_{1,\Lambda}=0.02$), where $T_g$ denotes the crossover temperature below which disorder becomes significant (see text). In the presence of the term $\tilde W_{3,k}$, $\tilde t_{k_x}$ is undefined for $T<T_c$ due to the flow singularity (see Sec.~\ref{subsubsec_Tc}).}
    \label{fig_tkx} 
\end{figure}

\begin{figure}
	\includegraphics[width=7.cm]{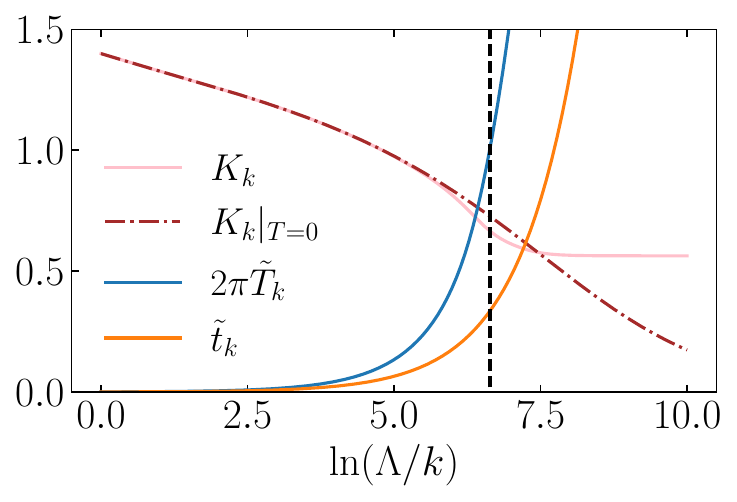}
	\hspace*{-0.67cm}
	\includegraphics[width=7.6cm]{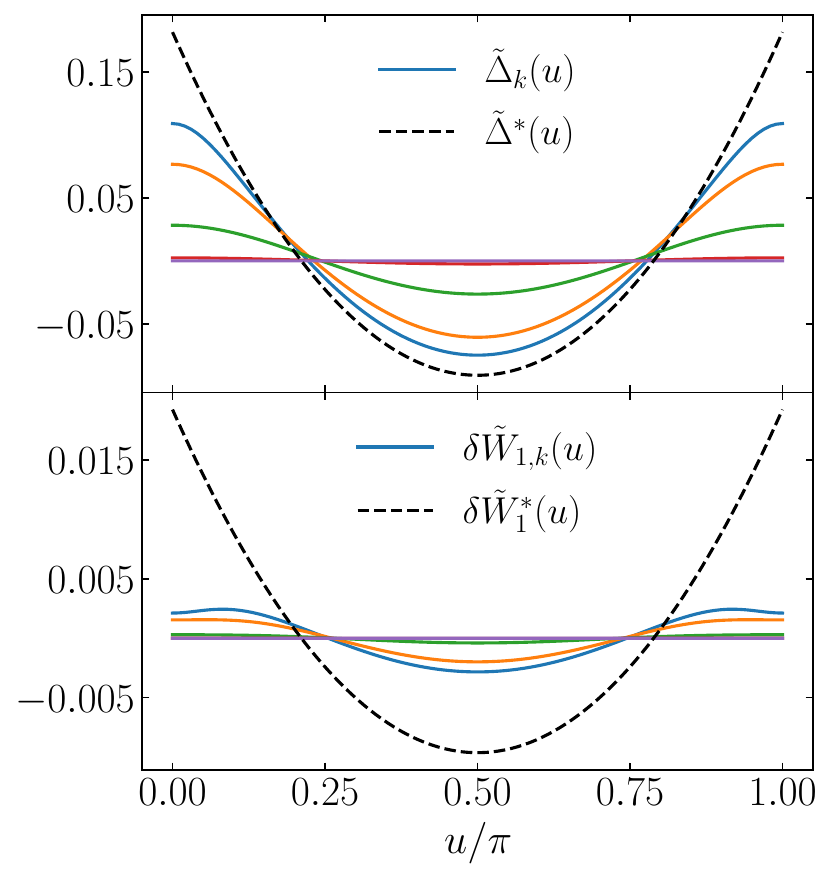}
\caption{RG flow at finite temperature for $\tilde W_{3,k}=0$ and $T/v\Lambda\simeq 10^{-4}$ ($K=1.4$, $\tilde \Delta_{1,\Lambda}=0.02$).  Top panel: $K_k$, $\tilde T_k$ and $\tilde t_k$ as a function of $\ln(\Lambda/k)$. The dash-dotted line shows $K_k|_{T=0}$ and the dotted vertical line marks the quantum-classical crossover scale $k_x$ defined by the criterion $\tilde T_{k_x}=1/2\pi$. Bottom panel: $\tilde\Delta_k(u)$ and $\delta\tilde W_{1,k}(u)$ for $k/k_x=0.419/0.225/0.120/0.064/0.035$ (from bottom to top for $u=\pi/2$).
	} 
	\label{fig_DeltaW_finiteT_1_withoutW3}
\end{figure}

\subsection{Flow without $\tilde W_{3,k}$}
\label{subsec_finiteT_withoutW3} 

In the absence of $\tilde W_{3,k}$, the quantum regime of the flow ($k\gg k_x$) is similar to the zero-temperature flow discussed in Sec.~\ref{subsec_Tzero_withoutW3}. The Luttinger parameter $K_k$ decreases, and the functions $\tilde\Delta_k$ and $\tilde W_{1,k}$ increase in amplitude. In this regime, temperature has negligible impact. 

To analyze the classical part of the flow ($k\lesssim k_x$), we distinguish two regimes according to the value of the dimensionless renormalized temperature 
\beq 
\tilde t_{k_x} = \pi K_{k_x} \tilde T_{k_x} = \frac{K_{k_x}}{2} 
\eeq  
at the quantum-classical crossover scale (Fig.~\ref{fig_tkx}). 

In the high-temperature regime, where $\tilde t_{k_x}$ is of order unity (though slightly below $1$), the functions $\tilde\Delta_k(u)$ and $\delta\tilde W_{1,k}(u)$ rapidly decrease in the classical part of the flow, and the flow trajectory approaches the normal-fluid fixed point ($\tilde t_k\to \infty,\tilde\Delta^*(u)=\tilde W_{1}^*(u)=0$) (Fig.~\ref{fig_DeltaW_finiteT_1_withoutW3}).  

\begin{figure}
	\includegraphics[width=7.cm]{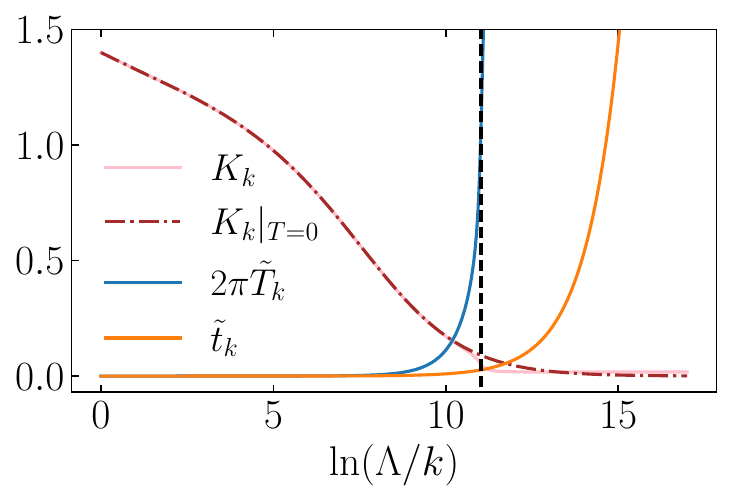}
	\hspace*{-0.66cm}
	\includegraphics[width=7.6cm]{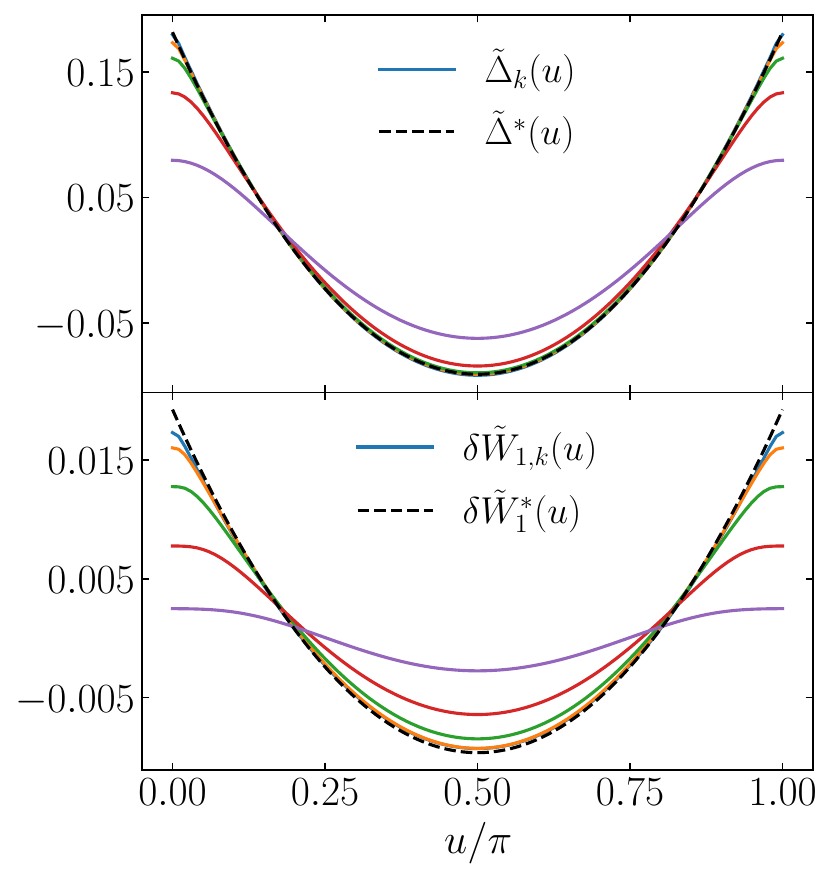}
	\caption{As in Fig.~\ref{fig_DeltaW_finiteT_1_withoutW3}, but for $T/v\Lambda\simeq 10^{-7}$ and $k/k_x=0.613/0.256/0.107/0.045/0.019$ (bottom panel, from bottom to top for $u=\pi/2$) . Unlike the high-temperature case shown in Fig.~\ref{fig_DeltaW_finiteT_1_withoutW3}, at the quantum-classical crossover, $\tilde t_{k_x}$ is much smaller than unity and $\tilde\Delta_{k_x}$ and $\delta\tilde W_{1,k_x}$ are close to their values at the localized fixed point~(\ref{fppot}) (shown as dashed curves).  } 
	\label{fig_DeltaW_finiteT_2_withoutW3}
\end{figure}

\begin{figure}
\centerline{\includegraphics[width=9cm]{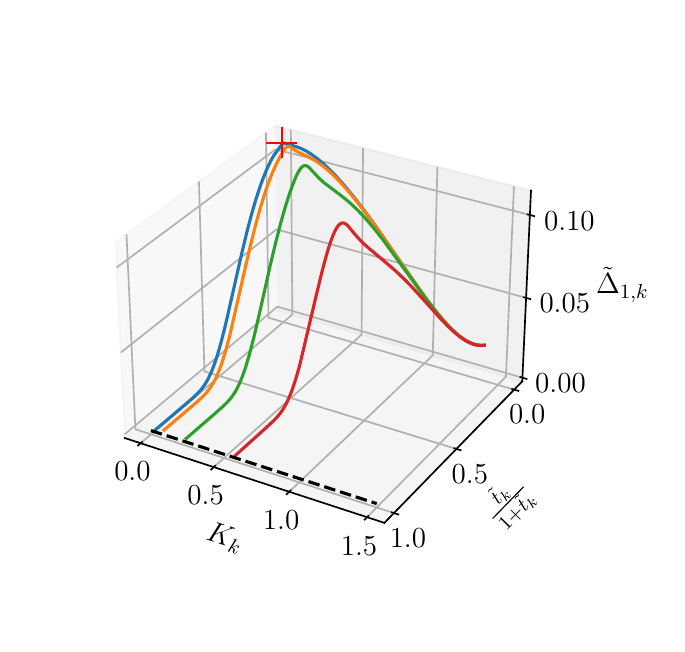}}
\caption{Flow diagram projected onto the space $(K_k,\tilde\Delta_{1,k},\tilde t_k)$ for $\tilde W_{3,k}=0$ and various temperatures ($K=1.4$, $\tilde \Delta_{1,\Lambda}=0.02$). The red cross indicates the localized fixed point, defined by $K_k=0$, or $\tilde t_k=0$, and~(\ref{fppot}). The dashed (black) line represents the normal-fluid fixed point. Since $K_k$ plays no role near this fixed point ---only the renormalized temperature $\tilde t_k$ matters--- this line should be viewed as a single fixed point. At low temperatures, the trajectories spend a significant amount of RG time near the localized fixed point, before eventually flowing toward the normal-fluid fixed point $(\tilde\Delta_1^*=0, \tilde t^*=\infty)$.}
\label{fig_flowdiag3D_withoutW3}
\end{figure}

In the low-temperature regime, below a characteristic temperature $T_g$, the renormalized coupling $\tilde t_{k_x}$ is much smaller than unity, and the behavior of the flow is significantly different (Fig.~\ref{fig_DeltaW_finiteT_2_withoutW3}). The functions $\tilde\Delta_{k_x}(u)$ and $\delta\tilde W_{1,k_x}(u)$ at the quantum-classical crossover are not very different from their values~(\ref{fppot}) at the localized fixed point. For $k\lesssim k_x$, the flow trajectory first moves closer to the localized fixed point, but eventually escapes from the fixed point when $\tilde t_k$ becomes of order unity. This leads us to define a thermal scale $k_T$ by
\begin{equation} 
	\tilde t_{k_T} = 1 , \quad \mbox{i.e.} \quad k_T = \frac{\pi TK}{v} .  
\end{equation}
The flow in the regime $k_T\lesssim k\lesssim k_x$ is classical and controlled by the zero-temperature ($\tilde t_k=0$) localized fixed point. Thus, we expect physical properties at intermediate length and time scales to be characteristic of a localized (glassy) phase with weak fluctuations. Only on larger scales do we expect to observe the characteristic properties of a thermal state dominated by thermal fluctuations. Several trajectories, corresponding to different temperatures, are shown in Fig.~\ref{fig_flowdiag3D_withoutW3}. 

\begin{figure}
\centerline{\includegraphics[width=7.5cm]{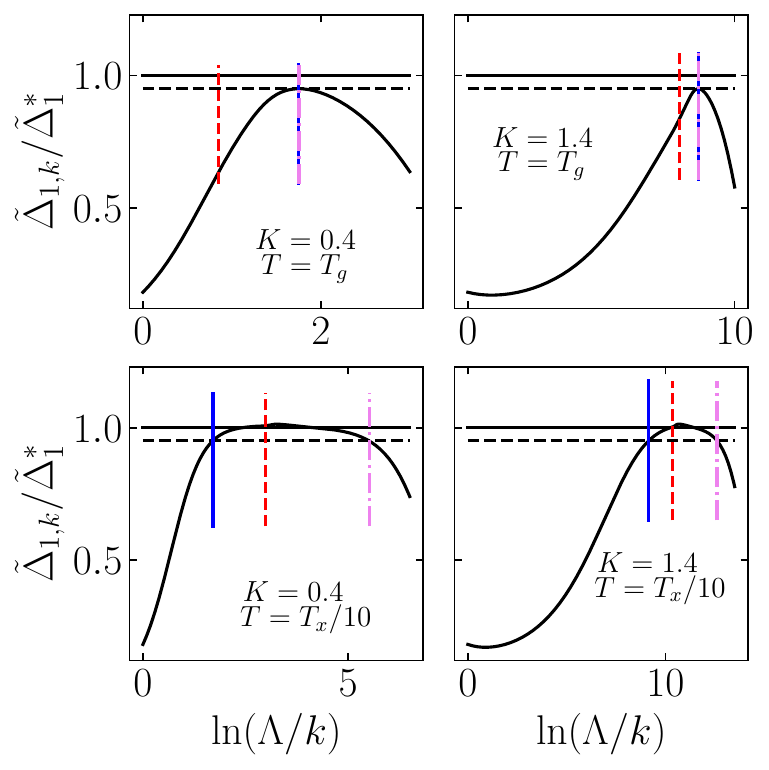}}
\caption{Ratio $\tilde\Delta_{1,k}/\tilde\Delta^*_1$ as a function of $\ln(\Lambda/k)$, for $K=0.4$ (left) and $1.4$ (right), and for $T=T_g$ (top) and $T=T_x/10$ (bottom). The solid (blue) vertical line indicates the scale $\kdis$, the dashed (red) line marks the scale $k_x$, and the dash-dotted (pink) line shows the scale $k_f$.}	
\label{fig_plateaux} 
\end{figure}

To determine the crossover temperature $T_g$, below which the flow trajectories are controlled by the localized fixed point at intermediate length and time scales, we adopt the following criterion. For temperatures below $T_g$, the amplitude $\tilde\Delta_{1,k}$ exceeds $\alpha \tilde\Delta^*_1$ before being suppressed as the trajectory flows toward the normal-fluid fixed point (we choose $\alpha=0.95$ in practice). This condition, $\tilde\Delta_{1,k}\geq \alpha \tilde\Delta^*_1$, defines a window $[\kdis,k_f]$. The disorder scale $\kdis\equiv \kdis(K,T)$ is the RG scale at which disorder effects become significant, while the final scale $k_f\equiv k_f(K,T)$ marks where thermal fluctuations start to suppress disorder. The temperature $T_g$ is thus defined by 
\beq 
\kdis(K,T_g) = k_f(K,T_g) . 
\label{Tgdef} 
\eeq
Figure~\ref{fig_plateaux} shows the flow of $\tilde\Delta_{1,k}$ for $T\ll T_g$ and at $T_g$, and at two values of the Luttinger parameter: $K=1.4$ and $K=0.4$. At low temperatures, the disorder scale $\kdis$ is much larger than $k_x$, whereas near $T_g$, the reverse holds: $k_x>\kdis$. To distinguish these two regimes, we define a second crossover temperature $T_x$ by 
\beq 
\kdis(K,T_x) = k_x(K,T_x) . 
\label{Txdef} 
\eeq

Thus, the normal-fluid phase is divided into three distinct regimes. For $T<T_x$, disorder effects are strong and quantum fluctuations dominate, as the flow from $k=\Lambda$ down to $\kdis\gg k_x$ is essentially insensitive to temperature. The disorder scale can be approximated by $\kdis(K,0)$, which serves as an estimate of the inverse of the zero-temperature localization length $\xiloc(K)$ in the Bose-glass phase~\cite{no4}. This regime can be viewed as a quantum glassy phase, where the correlation length of the density field $\varphi$ is expected to be of the order of $\xiloc(K)$. For $T_x<T<T_g$, disorder effects remain significant, but the disorder scale is now governed by thermal fluctuations, since $\kdis<k_x$. This defines a classical glassy regime, where the correlation length is expected to be of the order of the thermal length $v/T$. Finally, in the regime $T>T_g$, disorder is no longer relevant, and the flow trajectories do not approach the localized fixed point. The crossover temperatures $T_g$ and $T_x$, obtained from the numerical solution of the flow equations, are shown in Fig.~\ref{fig_TgTx_W30}.
  
One can obtain an approximate value of $T_x$ by approximating $\kdis(K,T_x)$ by $\kdis(K,0)=1/\xiloc(K)$, which yields 
\beq 
T_x \simeq \frac{v_{k_x}}{2\pi \xiloc(K)} \sim \frac{v}{2\pi \xiloc(K)} , 
\label{Txestimate1} 
\eeq
where the second expression neglects the renormalization of the velocity. Estimating $\xiloc(K)$ from the criterion $\tilde\Delta_{1,\kdis}=\alpha\tilde\Delta^*_1$ and the $T=0$ one-loop RG equation (with the renormalization of $K$ neglected), 
\beq 
\dt \tilde\Delta_{1,k} = - (3 - 2K) \tilde\Delta_{1,k} , \quad \frac{\tilde\Delta_{1,k}}{\tilde\Delta_{1,\Lambda}} = \left(\frac{\Lambda}{k}\right)^{3-2K} , 
\label{Delta1oneloop}
\eeq 
we obtain
\begin{align}
\xiloc(K) &= \frac{1}{\Lambda} \left( \frac{\alpha\tilde\Delta_1^*}{\tilde\Delta_{1,\Lambda}} \right)^{1/(3-2K)} \nonumber\\
&=  \frac{1}{\Lambda} \left( \frac{\alpha\tilde\Delta_1^* v^2\Lambda^3}{8\calD K^2} \right)^{1/(3-2K)} .
\label{xilocestimate} 
\end{align}
The localization length $\xiloc(K)$ has a well-defined limit as $K\to 0$, since $\tilde\Delta_{1,\Lambda}$ (i.e., $\calD K^2$) is held fixed. From Eqs.~(\ref{Txestimate1}) and (\ref{xilocestimate}), we find
\beq 
T_x 
\sim \frac{v\Lambda}{2\pi} \left( \frac{\tilde\Delta_{1,\Lambda}}{\alpha\tilde\Delta_1^*} \right)^{1/(3-2K)} .
\label{Txestimate2} 
\eeq
Neglecting the renormalization of the velocity in this estimate is consistent with he approximation used in the one-loop flow equation~(\ref{Delta1oneloop}). The estimate~(\ref{Txestimate2}) agrees qualitatively with the numerical result shown in Fig.~\ref{fig_TgTx_W30}: It predicts that $T_x$ increases upon decreasing $K$, and reaches a finite value as $K\to 0$.

In the regime $T\gg T_x$, the quantum-classical crossover occurs before disorder effects become significant, and the disorder scale $\kdis$ is primarily controlled by thermal fluctuations. One can then estimate $T_g$ from the classical flow equations~(\ref{eqcl}). To one-loop order, 
\beq 
\dt \tilde\Delta_{1,k} = -  \left( 3 - \frac{4}{\pi} \bar l_1(0) \tilde t_k \right) \tilde\Delta_{1,k} , 
\eeq 
i.e. 
\beq 
\tilde\Delta_{1,k} = \tilde\Delta_{1,\Lambda} \exp\left[ 3\ln\left(\frac{\Lambda}{k}\right) - \frac{4T\bar l_1(0)}{\pi Z_x k} \right] , 
\label{D1cloneloop} 
\eeq 
so that $\tilde\Delta_{1,k}$ reaches a maximum at $\kmax$, defined by $\bar l_1(0)\tilde t_{\kmax}=3\pi/4$. An explicit expression of the crossover temperature $T_g$, defined by the criterion $\tilde\Delta_{1,\kmax}=\alpha\tilde\Delta^*_1$ and $\kmax=k_f$ [Eq.~(\ref{Tgdef})], can be obtained by noting that $k_f\simeq k_T$ (the numerical solution of the flow equations gives $k_T/k_f\simeq 0.408$). This gives 
\beq 
T_g \sim \frac{3v\Lambda}{4K\bar l_1(0)}  \left( \frac{\tilde\Delta_{1,\Lambda}}{\alpha\tilde\Delta_1^* e^3} \right)^{1/3} .
\label{Tgestimate} 
\eeq 
The $1/K$ divergence as $K\to 0$ is consistent with the numerical result shown in Fig.~\ref{fig_TgTx_W30}. 

The finite-temperature crossover diagram has been studied by Glatz and Nattermann~\cite{Glatz02,Glatz04} using the one-loop RG equations. In the normal-fluid phase, by comparing characteristic length scales, they identified three distinct regimes that are similar to the three regimes discussed above. However, the importance of the zero-temperature localized fixed point for the finite-temperature properties cannot be fully understood from the perturbative approach.  

\subsection{Flow with $\tilde W_{3,k}$}

\subsubsection{Flow in the normal-fluid phase} 
\label{subsec_normalfluid}

At sufficiently high temperatures, the flow remains regular and can be integrated from $k=\Lambda$ down to $k=0$. Its behavior closely resembles the case without $\tilde W_{3,k}$ (Fig.~\ref{fig_flowdiag3D_withoutW3}). As before, two distinct regimes can be identified based on the value of the renormalized temperature $\tilde t_{k_x}$ at the quantum-classical crossover scale (Fig.~\ref{fig_tkx}). In the high-temperature regime, where $\tilde t_{k_x}\sim 1$, the flow trajectories are rapidly attracted to the normal-fluid fixed point, and the localized fixed point becomes irrelevant. By contrast, for temperatures below the crossover scale $T_g$, the renormalized temperature $\tilde t_{k_x}$ is much smaller than unity, and the RG trajectories spend a significant amount of time near the localized fixed point before ultimately being drawn to the normal-fluid fixed point. 

\subsubsection{Finite-temperature phase transition} 
\label{subsubsec_Tc}
At low temperatures, a singularity appears in the flow at a finite RG scale $k_c$, similar to the behavior in the zero-temperature limit. This scale $k_c$ is slightly smaller but remains of the same order as the inverse zero-temperature localization length $\kdis(K,0)=1/\xiloc(K)$. As the temperature increases, the singularity is suppressed due to the quantum-classical crossover when $k_x$ becomes comparable to $k_c\sim 1/\xiloc(K)$, corresponding to $T\gtrsim T_x$. We therefore expect the singularity to occur below a temperature $T_c$, which can be estimated by the condition $k_x\sim k_c \sim 1/\xiloc(K)$, yielding 
\beq 
T_c \sim \frac{v_{k_x}}{2\pi \xiloc(K)} \sim  \frac{v}{2\pi \xiloc(K)} .
\label{Tcestimate} 
\eeq
This estimate coincides with the crossover temperature $T_x$ in the absence of the term $\tilde W_{3,k}$ [Eq.~(\ref{Txestimate2})]. Thus, we conclude that the term $\tilde W_{3,k}$ has little effect in the classical glassy regime ($T_x\lesssim T\lesssim T_g$), but destabilizes the quantum glassy regime ($T\lesssim T_x$).  

Figure~\ref{fig_TcEstim} shows $T_c$ as a function of $K$, together with the crossover temperatures $T_g$ and $T_x$ obtained from the conditions~(\ref{Tgdef}) and (\ref{Txdef}). Apart from an unimportant prefactor, the estimate~(\ref{Tcestimate}) proves to be accurate. The crossover temperature $T_x$, evaluated with the $\tilde W_{3,k}$ term, remains comparable to $T_c$, as expected from the qualitative argument above. Nevertheless, for $K\gtrsim 0.3$, $T_x$ is slightly larger than $T_c$, giving rise to a quantum glassy regime over a narrow temperature range, while it becomes undefined for $K\lesssim 0.3$.   

Our findings are reminiscent of the work of MAAS, who argued for the existence of a finite-temperature localization-delocalization transition~\cite{Michal16}. Their estimate of the transition temperature is consistent with the expression~(\ref{Tcestimate}) for the temperature $T_c$ below which a singularity in the flow arises. This suggests that the singularity observed in the FRG flow is not a mere artifact, but rather signals a genuine finite-temperature fluid-insulator transition.

\section{Relation to MBL}
\label{sec_mbl} 

The concordance between our findings (with the inclusion of the term $\tilde W_{3,k}$) and those of MAAS prompts us to analyze our results from the MBL point of view~\cite{Alet18,Abanin19,Sierant25}. Originally, MBL was presented as a finite-temperature phase transition between an insulating state and a conducting state in an interacting quantum many-body system~\cite{Basko06,Gornyi05}. Subsequent works focused on high-energy states and defined MBL as a nonergodic phase of matter that avoids thermalization~\cite{Oganesyan07,Pal10,Nandkishore15}. Although these studies deal with highly nonequilibrium states with finite energy density, we will see that there are analogies with the low-temperature one-dimensional Bose gas studied in this manuscript~\cite{Dupont2019}. Before turning to this discussion, we stress that the existence of MBL as a true phase of matter remains an open question~\cite{Suntajs20}, even though many systems show no sign of thermalization at finite size and over finite evolution times~\cite{Sierant25}. In this section, we consider only the results obtained when the term $\tilde W_{3,k}$ is included.

\subsection{Lack of thermalization} 

The defining property of MBL is the absence of thermalization (i.e., the breaking of ergodicity)~\cite{Alet18,Abanin19,Sierant25}, which implies a breakdown of the eigenstate thermalization hypothesis~\cite{Deutsch91,Srednicki94,Rigol08}. A consequence of ergodicity breaking is the inapplicability of standard methods of equilibrium statistical mechanics. In particular, finite-temperature field-theory techniques based on the (Euclidean) Matsubara formalism, which assume that the density operator is given by the Boltzmann-Gibbs distribution, are therefore inadequate for studying the MBL phase. 

In the MBL scenario, the finite-scale singularity ---beyond which the flow cannot be integrated---  that arises below the temperature $T_c$ would be interpreted as a signature of nonergodicity. In the following section, we discuss how this can be understood within the framework of the droplet picture introduced in Sec.~\ref{subsec_Tzero_withW3}.

\subsection{Avalanche mechanism vs droplet picture}
\label{subsec_avalanches} 

According to the quantum-avalanche mechanism, the MBL transition is due to the growth of small ergodic (i.e., thermal) bubbles that destabilize the surrounding MBL regions~\cite{DeRoeck17,Thiery18,Goremykina19}. In a disordered system, rare regions of anomalously weak disorder and arbitrary size exist in the thermodynamic limit with probability one, and are likely to be ergodic. When the localization length $\xi$ of the MBL regions exceeds a critical value $\xi^*$, these rare ergodic inclusions can thermalize the whole system through a mechanism of quantum avalanches. The critical point of the MBL transition is a smooth continuation of the MBL phase, and the localization length $\xi<\xi^*$ does not diverge at the transition. Nevertheless, it has been proposed that a characteristic length $\ell^*\sim 1/(\xi^*-\xi)$, at which the power-law distribution of thermal bubbles is cut off by an exponential tail, diverges at the MBL transition~\cite{Thiery18,[{See also }]Laflorencie20}.

In Sec.~\ref{subsec_Tzero_withW3}, the FRG flow in the one-dimensional Bose gas was interpreted within a droplet picture. In Ref.~\cite{Dupuis20}, it was proposed that quantum-mechanically active droplets, which couple to the ground state {\it via} quantum tunneling, give rise to ``superfluid'' domains with significant density fluctuations and, consequently, suppressed phase fluctuations of the boson field. In this scenario, droplet excitations allow the system to thermalize locally and are reminiscent of thermal bubbles in MBL phases. In Sec.~\ref{subsec_Tzero_withW3}, we argued that the low-temperature localized phase in the one-dimensional Bose gas arises from the absence of quantum-mechanically active droplets larger than a characteristic size $\xi_c\sim 1/k_c$ (comparable to the zero-temperature localization length, see Sec.~\ref{subsubsec_Tc}). As a consequence, the system is unable to thermalize on scales larger than $\xi_c$. The distribution of quantum-mechanically active droplets is expected to follow a power law, consistent with the standard droplet picture.

Alternatively, droplets can be viewed as signatures of many-body resonances (MBR). Within the picture described in Sec.~\ref{subsec_Tzero_withoutW3}, these resonances arise between two configurations of the field $\varphi$ that differ by a soliton-antisoliton pair, i.e., by a particle-hole excitation. This is reminiscent of the MBR scenario proposed for MBL systems~\cite{Sierant25}, where, in the simplest picture, a resonance involves only two states and take the form of cat states constructed from their two independent linear combinations~\cite{Villalonga20,Villalonga20a,Crowley22,Laflorencie25}. If MBR exist on all scales, the system is thermal. Conversely, the system remains in the localized phase if the size of MBR is bounded by $\xi_c\sim 1/k_c$.

\subsection{Many-body spectrum} 

The many-body spectrum of a system varies with the disorder strength $W$. In the small-disorder limit, the system is ergodic, and the evolution of the eigenenergies $E_n(W)$ with $W$ is characterized by avoided level crossings, as expected for chaotic systems~\cite{Zakrzewski93a,Zakrzewski93,Zakrzewski23}. In contrast, for a system in the MBL phase (i.e., in the large-$W$ limit), energy levels exhibit exact crossings~\cite{Maksymov19}. These crossings can be seen as a manifestation of the existence of quasilocal integrals of motions (LIOMs) or $\ell$-bits~\cite{Huse14,Serbyn13}. These conserved quantities strongly constrain the system's dynamics, thereby providing an intuitive explanation for the absence of thermalization.

In the one-dimensional Bose gas, we expect the evolution of the spectrum with disorder strength to be determined by the presence or absence of quantum-mechanically active droplets. In Sec.~\ref{sec_zeroT}, we pointed out that quantum tunneling between the ground state and droplet excitations leads to avoided level crossings as an external source (or the disorder strength) is varied. Therefore, the absence of quantum-mechanically active droplets beyond a certain length scale is expected to result in level crossings when the disorder strength changes. This suggests a correspondence between the existence of LIOMs in MBL systems and the absence of quantum-mechanically active droplets beyond a characteristic length scale in the one-dimensional Bose gas. In both cases, the localization length plays a fundamental role, as it determines the range of the LIOMs, which are quasilocal operators, and sets the maximum size of quantum-mechanically active droplets. At length scales larger than the localization length, the LIOMs can be viewed as fully local, and one enters the MBL phase in which thermalisation becomes impossible.

\subsection{Slow dynamics in the ergodic phase} 

The MBL phase is characterized by extremely slow dynamics at finite times, regardless of whether this dynamics ultimately leads to thermalization in the infinite-time limit. If thermalization occurs, it is more accurate to refer to an MBL regime, rather than an MBL phase~\cite{Morningstar22,Sierant25}. Numerous signatures of slow dynamics have been observed in numerical studies (necessarily restricted to finite-time simulations) of the (putative) MBL phase, particularly in the context of the disordered XXZ spin chain: memory of the initial state~\cite{TorresHerrera14,TorresHerrera15,TorresHerrera18}, logarithmic growth of entanglement entropy~\cite{DeChiara06,Znidaric08,Bardarson12,Serbyn13a,Andraschko14}, etc. Some of these features persist, at least qualitatively, in the ergodic phase~\cite{Luitz17a}.

In the one-dimensional Bose gas, when the temperature is just above the transition temperature $T_c$, we find that the FRG flow is governed by a zero-temperature localized fixed point on intermediate scales, although it is eventually attracted to the normal-fluid fixed point. As discussed in the preceding sections, this implies that physical properties on intermediate length and time scales closely resemble those of a localized (glassy) phase with negligible thermal fluctuations. This bears a strong resemblance to an MBL regime, characterized by very slow dynamics at intermediate scales and thermalization in the long-time, long-distance regime. Notably, this MBL-like regime also appears in the absence of the term $\tilde W_{3,k}$ (Sec.~\ref{subsec_finiteT_withoutW3}).

\subsection{Critical behavior}

The FRG flow in the localized phase ($T<T_c$) is essentially independent of temperature, and the scale $k_c(T)$ at which the singularity occurs does not vanish, so the localization length $\xi_c(T)\sim 1/k_c(T)$ remains finite as $T\to T_c^-$. Furthermore, the divergence of the (running) dynamical critical exponent $z_k=\theta_k+1$ as $k\to k_c(T)$ bears similarities with the predicted divergence of the dynamical critical exponent at the MBL transition~\cite{Potter15,Vosk15}.

\section{Conclusion} 

We have studied the finite-temperature phase diagram of a one-dimensional disordered Bose gas within two different scenarios, obtained from two distinct truncations of the effective action. Both scenarios reveal a remarkable property: At low temperatures, below a crossover temperature $T_g$, the FRG flow is governed on intermediate scales by a zero-temperature localized fixed point. Consequently, we expect the physical properties on these scales to reflect those of a localized glassy phase. One can further distinguish a quantum glassy regime, where the correlation length is set by the zero-temperature localization length, from a classical glassy regime at higher temperatures, where the correlation length is set by the thermal length (see Figs.~\ref{fig_TgTx_W30} and \ref{fig_TcEstim}). This glassiness is reminiscent of the MBL regime (as opposed to an MBL phase) discussed in the context of MBL~\cite{Morningstar22,Sierant25}. 

Both scenarios are based on a derivative expansion of the effective action that includes the one- and two-replica terms (neglecting higher-order replica terms). The first scenario is obtained by retaining all second derivatives except the second time derivative in the two-replica term. It predicts that the Bose glass is unstable toward a normal-fluid phase at any finite temperature, in agreement with the predictions of perturbative RG~\cite{Glatz02,Glatz04}. 
	
The second scenario, obtained by retaining all second derivatives in the effective action, is based on interpreting a finite-scale singularity in the RG flow as the signature of a localized phase. This interpretation is consistent with the phase diagram proposed by MAAS, namely the existence of a finite-temperature fluid-insulator transition. We have explained the low-temperature insulating phase in a droplet picture, and highlighted its analogies with several key features of MBL phases. Although MBL is a property of highly excited states, it is natural to ask whether the existence of a Bose glass necessarily implies MBL at low temperatures~\cite{Abanin19}. 

We cannot exclude the possibility that the singularity in the RG flow is an artifact of the derivative expansion of the effective action used to solve the exact functional flow equation. However, improvements to the derivative expansion, such as the LPA$''$ or the Blaizot--Méndez-Galain--Wschebor approximation (Appendix~\ref{app_beyondDE}), do not suppress the singularity. It would be interesting to investigate the stability of our results with respect to the inclusion of a three-replica term in the effective action.

\section*{Acknowledgment}
We thank J. Colbois, N. Laflorencie, G. Lemari\'e, G. Tarjus, M. Tarzia, and M. Tissier for discussions and comments on the manuscript.

\appendix

\section{Statistical tilt symmetry} 
\label{app:STS} 

The STS is a property of the replicated action \eqref{Srep}. It is a consequence of the transformation of the action under a shift of the field, $\varphi_a(x,\tau)\to\varphi'_a(x,\tau)=\varphi_a(x,\tau)+w(x)$:
\begin{align}
    S[\{\varphi_a\}] = S[\{\varphi_a'\}] - \frac{n}{2} \beta (Z_x - 2n\beta\calF) \int_x (\dx w)^2 \nonumber \\
+ (Z_x - 2n\beta\calF) \int_{x,\tau} \sum_a \varphi_a(\dx^2 w) , 
\end{align}
where $Z_x=v/\pi K$. The partition function \eqref{Zrep} can be recast as
\begin{equation}
    \calZ[\{J_a\}] = \calZ[\{J_a'\}]e^{\frac{n}{2} \beta (Z_x - 2n\beta\calF) \int_x (\dx w)^2 - \int_{x,\tau}\sum_a J_a'w} ,
\end{equation}
with
\begin{equation}
    J_a'(x,\tau) = J_a(x,\tau) - (Z_x - 2n\beta\calF) \dx^2w .
\end{equation}
The associated effective action reads
\begin{align}
        \Gamma[\{\phi_a\}] = & -\ln \calZ[\{J_a\}]+\int_{x,\tau}\sum_a J_a\phi_a \nonumber \\
        = & -\ln \calZ[\{J_a'\}]+\int_{x,\tau}\sum_a J_a'\phi_a' \nonumber \\
        & - \frac{n}{2}\beta(Z_x - 2n\beta\calF)\int_x (\dx w)^2 \nonumber \\
        & + (Z_x - 2n\beta\calF) \int_{x,\tau} \sum_a \phi_a \dx^2 w ,
\end{align}
where we have introduced
\begin{align}
    & \phi_a(x,\tau) = \frac{\delta\ln \calZ[\{J_f\}]}{\delta J_a(x,\tau)} = \phi_a'(x,\tau) - w(x) \nonumber , \\
    & \phi_a'(x,\tau) = \frac{\delta\ln \calZ[\{J_f'\}]}{\delta J_a'(x,\tau)} .
\end{align}
This leads to  
\begin{align}
    \Gamma[\{\phi_a\}] ={}& \Gamma[\{\phi_a+w\}] - \frac{n}{2} \beta (Z_x - 2n\beta\calF) \int_x (\dx w)^2 \nonumber\\ & 
- (Z_x - 2n\beta\calF) \int_{x,\tau} \sum_a (\dx w)(\dx\phi_a) . 
\label{STSgamma}
\end{align}
From \eqref{STSgamma} and the arbitrariness of the number $n$ of replicas we deduce that i) the effective action $\Gamma_k[\{\phi_a\}]$ is invariant under a constant (that is, space- and time-independent) shift of the field; ii) the only possible term with spatial derivatives in $\Gamma_1[\phi_a]$ is the one present in the bare action, and its coefficient $Z_x$ is therefore not renormalized; iii) similarly, the only terms involving purely spatial derivatives in $\Gamma_2$ are second order in $\dx$ and must sum to $2\calF$; iv) all higher-order cumulants are invariant under the shift $\phi_a(x,\tau)\to \phi_a(x,\tau)+w(x)$. It is important to note that we did not consider the effect of the regulator in this proof. However, we can show in a similar fashion that the subtraction of the regulator term in the definition \eqref{modifiedGamma} of the effective action makes the additional terms disappear, so that all the properties just listed also hold for the scale-dependent effective action $\Gamma_k[\{\phi_a\}]$.

\section{Flow equations} 
\label{app:floweq} 

We give below the complete flow equations, obtained with the ansatz~(\ref{ansatz1},\ref{ansatz2}):
\begin{widetext}
\begin{equation}
    \theta_k = \frac{\pi}{2} \, \bigl[ \bar M^\tau_{0,1}(0)   \tilde\Delta_k''(0)   +   \bar M^\tau_{0,1}(1)   \tilde{W}_{1,k}''(0)   -   \bar l_1(0)   \tilde{W}_{3,k}''(0)   -   8   \tilde\Omega^2   \bar M^\tau_{0,1}(0)   \tilde{W}_{3,k}''(0) \bigr] , 
\end{equation}
\begin{align}
       \partial_t \tilde\Delta_k(u) = & -3 \tilde\Delta_k (u) -K_k l_1(0,0) \tilde\Delta_k ''(u) + K_k l_1(1,0) \tilde{W}_{2,k}''(u) + K_k l_1(0,1) \tilde{W}_{3,k}''(u) + \pi \bar l_2(0) [\tilde\Delta_k (u) -\tilde\Delta_k (0) ]  \tilde\Delta_k ''(u) \nonumber\\
   & + \pi  \bar l_2(0) \tilde\Delta_k '(u)^2 + \pi  \bar l_2(1) [ \tilde{W}_{1,k}(u) - \tilde{W}_{1,k}(0) ] \tilde\Delta_k ''(u) + 2 \pi  \bar l_2(1) \tilde\Delta_k '(u) \tilde{W}_{1,k}'(u) +\pi  \bar l_2(2) \tilde{W}_{1,k}'(u){}^2 \nonumber\\
   & + \pi  \bar l_2(1) \tilde\Delta_k (u) \tilde{W}_{1,k}''(u) +\pi \bar l_2(1) \tilde\Delta_k (0)  \tilde{W}_{2,k}''(u)+\pi  \bar l_2(2)
   \tilde{W}_{1,k}(u) \tilde{W}_{1,k}''(u)+\pi  \bar l_2(2) \tilde{W}_{1,k}(0) \tilde{W}_{2,k}''(u) , 
\end{align}
\begin{align}
        \partial_t \tilde W_{1,k}(u) = & -\tilde{W}_{1,k}(u) -K_k l_1(0,0) \tilde{W}_{1,k}''(u) + \pi  \bar m^x_{1,1}(0) \tilde\Delta_k '(u)^2 + 2 \pi  \bar m^x_{1,1}(1) \tilde\Delta_k '(u) \tilde{W}_{1,k}'(u) + \pi \bar m^x_{1,1}(2) \tilde{W}_{1,k}'(u){}^2  \nonumber\\
   &  + \pi  \bar l_2(0) [ \tilde\Delta_k (u) - \tilde\Delta_k (0) ] \tilde{W}_{1,k}''(u) + \pi  \bar l_2(1) [ \tilde{W}_{1,k}(u) - \tilde{W}_{1,k}(0)  ] \tilde{W}_{1,k}''(u) - \pi  \bar l_2(0) \tilde\Delta_k '(u) \tilde{W}_{2,k}'(u)   \nonumber\\
   & + \pi \bar l_2(1) [ \tilde{W}_{1,k}'(u){}^2 - \tilde{W}_{2,k}'(u){}^2 ] - 3 \pi  \bar l_2(1) \tilde{W}_{1,k}'(u) [ \tilde{W}_{1,k}'(u)+ \tilde{W}_{2,k}'(u) ] ,
\end{align}
\begin{align}
        \partial_t \tilde{W}_{2,k}(u) = & -\tilde{W}_{2,k}(u) -K_k l_1(0,0) \tilde{W}_{2,k}''(u) - \pi  \bar m^x_{1,1}(0) \tilde\Delta_k '(u)^2 - 2 \pi  \bar m^x_{1,1}(1) \tilde\Delta_k '(u) \tilde{W}_{1,k}'(u) - \pi \bar m^x_{1,1}(2) \tilde{W}_{1,k}'(u){}^2\nonumber \\
   &  + \pi  \bar l_2(0) [ \tilde\Delta_k (u) - \tilde\Delta_k (0) ] \tilde{W}_{2,k}''(u) + \pi  \bar l_2(1) [ \tilde{W}_{1,k}(u) - \tilde{W}_{1,k}(0) ] \tilde{W}_{2,k}''(u) + \pi  \bar l_2(0) \tilde\Delta_k '(u) \tilde{W}_{2,k}'(u)  \nonumber \\
   & - \pi  \bar l_2(0) \tilde\Delta_k (u) [ \tilde{W}_{1,k}''(0) +  \tilde{W}_{2,k}''(0) ] + \pi  \bar l_2(1) \tilde{W}_{2,k}(u) [ \tilde{W}_{1,k}''(0) + \tilde{W}_{2,k}''(0) ] \nonumber\\ & - \pi  \bar l_2(1) \tilde{W}_{2,k}'(u) [ \tilde{W}_{1,k}'(u) + \tilde{W}_{2,k}'(u) ] ,
\end{align}
\begin{align}
        \partial_t \tilde{W}_{3,k}(u) = & - 3 \tilde{W}_{3,k}(u) + 2 z \, \tilde{W}_{3,k}(u) -K_k   l_1(0,0)   \tilde{W}_{3,k}''(u)
        -2 \pi   \bar M^\tau_{1,1}(0)   \tilde\Delta_k''(0)   \tilde\Delta_k(u)   + 2\pi   \bar M^\tau_{1,1}(1)   \tilde\Delta_k''(0) \tilde{W}_{2,k}(u) \nonumber\\& -2 \pi   \bar M^\tau_{1,1}(1)   \tilde\Delta_k(u)   \tilde{W}_{1,k}''(0)  + 2 \pi   \bar M^\tau_{1,1}(2)   \tilde{W}_{2,k}(u)   \tilde{W}_{1,k}''(0) 
   -\pi   \bar M^\tau_{1,1}(0)   \tilde\Delta_k'(u)^2 -2 \pi   \bar M^\tau_{1,1}(1)   \tilde\Delta_k'(u)   \tilde{W}_{1,k}'(u)\nonumber\\&  -\pi  \bar M^\tau_{1,1}(2)   \tilde{W}_{1,k}'(u)^2
    + \pi   \bar l_2(0)   \tilde\Delta_k''(0)   \tilde{W}_{3,k}(u) + \pi   \bar l_2(1)   \tilde{W}_{1,k}''(0)   \tilde{W}_{3,k}(u) + \pi   \bar l_2(0)   \tilde\Delta_k (u)   \tilde{W}_{3,k}''(0)  \nonumber\\& -\pi   \bar l_2(1)   \tilde{W}_{2,k}(u)   \tilde{W}_{3,k}''(0) 
   + \pi   \bar l_2(0)   [ \tilde\Delta_k(u) - \tilde\Delta_k(0) ]   \tilde{W}_{3,k}''(u) + \pi   \bar l_2(1)   [ \tilde{W}_{1,k}(u) - \tilde{W}_{1,k}(0) ]  \tilde{W}_{3,k}''(u) \nonumber\\& + 2 \pi   \bar l_2(0)   \tilde\Delta_k'(u)   \tilde{W}_{3,k}'(u) 
   + 2 \pi   \bar l_2(1)   \tilde{W}_{1,k}'(u)   \tilde{W}_{3,k}'(u)  
   + \bigl[ -2 \pi   \bar M^\tau_{0,2}(0)   \tilde\Delta_k (u)   \tilde\Delta_k''(0) + 2 \pi  
   \bar M^\tau_{0,2}(1)   \tilde{W}_{2,k}(u)   \tilde\Delta_k ''(0) \nonumber\\& + 8 \pi   \bar M^\tau_{1,1}(0)   \tilde{W}_{3,k}(u)   \tilde\Delta_k''(0) 
   -2 \pi   \bar M^\tau_{0,2}(1)   \tilde\Delta_k(u)   \tilde{W}_{1,k}''(0) + 2 \pi   \bar M^\tau_{0,2}(2)   \tilde{W}_{2,k}(u)   \tilde{W}_{1,k}''(0) \nonumber \\& + 8 \pi   \bar M^\tau_{1,1}(1)   \tilde{W}_{3,k}(u)   \tilde{W}_{1,k}''(0)  
   + 16 \pi   \bar M^\tau_{1,1}(0)   \tilde\Delta_k(u)   \tilde{W}_{3,k}''(0) -16 \pi   \bar M^\tau_{1,1}(1)   \tilde{W}_{2,k}(u)   \tilde{W}_{3,k}''(0) \nonumber \\& -8 \pi   \bar l_2(0)   \tilde{W}_{3,k}(u)   \tilde{W}_{3,k}''(0) 
   + 8 \pi   \bar M^\tau_{1,1}(0)   \tilde\Delta_k'(u)   \tilde{W}_{3,k}'(u) + 8 \pi   \bar M^\tau_{1,1}(1)   \tilde{W}_{1,k}'(u)   \tilde{W}_{3,k}'(u) -4 \pi   \bar l_2(0)   \tilde{W}_{3,k}'(u)^2 \bigr]   \tilde\Omega^2 \nonumber\\
    & + \bigl[ -16 \pi   \bar M^\tau_{1,1}(0)   \tilde{W}_{3,k}'(u)^2 + 8 \pi   \bar M^\tau_{0,2}(0)   \tilde{W}_{3,k}(u)   \tilde\Delta_k''(0) + 8 \pi   \bar M^\tau_{0,2}(1)   \tilde{W}_{3,k}(u)   \tilde{W}_{1,k}''(0) \nonumber \\
   & + 16 \pi   \bar M^\tau_{0,2}(0)   \tilde\Delta_k(u)   \tilde{W}_{3,k}''(0) -16 \pi   \bar M^\tau_{0,2}(1)   \tilde{W}_{2,k}(u)   \tilde{W}_{3,k}''(0) -64 \pi   \bar M^\tau_{1,1}(0)   \tilde{W}_{3,k}(u)   \tilde{W}_{3,k}''(0) \bigr]   \tilde\Omega^4\nonumber \\
   &  -64 \pi \bar M^\tau_{0,2}(0) \tilde{W}_{3,k}(u) \tilde{W}_{3,k}''(0)  \tilde\Omega^6 , 
    \label{W3eq}
\end{align}
where $\tilde\Omega=2\pi\tilde T_k = \tilde\omega_1$ is the first (dimensionless) Matsubara frequency, and the various threshold functions are defined in the next section.

\subsection{Threshold functions} 
\label{app:threshold} 

The threshold functions are defined by 
\begin{equation} 
\begin{split} 
    l_1(a,b) ={}& \tilde T_k \sum_{\twn} \int_0^\infty d\tilde q \, \tilde q^{2a} \tilde \omega_n^{2b} \frac{\partial_t R_k(\tilde q,i\tilde\omega_n)}{Z_x k^2} \tilde P_k(\tilde q,i\tilde\omega_n)^2 , \\ 
    \bar l_2(a) ={}& 2 \int_0^{+\infty}d\tilde q \,\tilde q^{2a} \frac{\partial_t R_k(\tilde q,0)}{Z_x k^2} \tilde P_k(\tilde q,0)^3 , \\ 
    \bar m^x_{1,1}(a) ={}& \frac{1}{2}\int_0^{+\infty}d\tilde q \, \tilde q^{2a} \biggl\{ \frac{\partial_t R_k(\tilde q,0)}{Z_x k^2} \tilde P_k(\tilde q,0)^2 \frac{\partial^2 \tilde P_k(\tilde q,0)}{\partial \tilde q^2}  + \tilde P_k(\tilde q,0) \frac{\partial^2}{\partial \tilde q^2}\left[ \frac{\partial_t R_k(\tilde q,0)}{Z_x k^2} \tilde P_k(\tilde q,0)^2\right] \biggr\} , 
\end{split} 
\end{equation} 
and 
\begin{equation}
\begin{split} 
\bar M^\tau_{0,1}(a) & = \frac{1}{2}\int_0^{+\infty}d\tilde q \, \tilde q^{2a}   \frac{\Delta^2}{\Delta \tilde \omega_n^2}\left[ \frac{\partial_t R_k(\tilde q,i\tilde\omega_n)}{Z_x k^2} \tilde P_k(\tilde q,i\tilde\omega_n)^2\right]\Bigg|_{\tilde\omega=0} , \\
\bar M^\tau_{1,1}(a) & = \frac{1}{2}\int_0^{+\infty}d\tilde q \, \tilde q^{2a} \biggl\{ \frac{\partial_t R_k(\tilde q,0)}{Z_x k^2} \tilde P_k(\tilde q,0)^2 \frac{\Delta^2 \tilde P_k(\tilde q,i\tilde\omega_n)}{\Delta \tilde \omega_n^2} + \tilde P_k(\tilde q,0) \frac{\Delta^2}{\Delta \tilde \omega_n^2}\biggl[ \frac{\partial_t R_k(\tilde q,i\tilde\omega_n)}{Z_x k^2} \tilde P_k(\tilde q,i\tilde\omega_n)^2\biggr] \biggr\} \bigg|_{\twn=0} , \\
\bar M^\tau_{0,2}(a) & = \int_0^{+\infty}d\tilde q \, \tilde q^{2a} \biggl\{ \frac{\Delta^2 \tilde P_k(\tilde q,i\tilde\omega_n)}{\Delta \tilde \omega_n^2} \frac{\Delta^2}{\Delta \tilde \omega_n^2}\left[ \frac{\partial_t R_k(\tilde q,i\tilde\omega_n)}{Z_x k^2} \tilde P_k(\tilde q,i\tilde\omega_n)^2\right] \biggr\} \bigg|_{\twn=0} ,
\end{split}
\end{equation} 
\end{widetext}
where 
\begin{align}
    \tilde P_k(\tilde q,i\tilde\omega_n) &= \frac{Z_x k^2}{\Gamma_{1,k}^{(2)}(q,i\omega_n)+R_k(q,i\omega_n)} \nonumber \\
    &= \frac{1}{(\tilde q^2 + \tilde \omega_n^2)[1+r(\tilde q^2 + \tilde \omega_n^2)]} ,
\end{align}
is the dimensionless propagator. The function $r(x)$ is defined after \eqref{regdef}. We use the notation 
\begin{equation}
    \frac{\Delta f(\tilde\omega_n)}{\Delta\tilde\omega_n} = \frac{f(\tilde\omega_n+\tilde\Omega)-f(\tilde\omega_n-\tilde\Omega)}{2\tilde\Omega} 
\end{equation}
for the finite differences in the definition of the functions $\bar M^\tau$. At zero temperature, where $\tilde\omega_n$ becomes a continuous variable, the Matsubara sums become frequency integrals, and the finite differences reduce to ordinary derivatives.

\subsection{Classical limit} 
\label{app:classical_limit} 

In the classical limit $\tilde T_k\gg 1$, all nonzero dimensionless Matsubara frequencies become much larger than unity, and only the component $\tilde\omega_{n=0}=0$ contributes significantly to the threshold functions. The contribution of nonzero Matsubara frequencies is suppressed by the regulator function. This indicates that the system behaves as a classical system in $d = 1$ spatial dimension, with the field being nearly time independent. In this limit, the threshold functions $l_1(a,b)$ with $b>0$ can be ignored, while $l_1(a,0)$ behaves as
\begin{equation}
     l_1(a,0) \underset{\tilde T_k \gg 1}{\simeq} \tilde T_k \bar l_1(a) ,
\end{equation}
with
\begin{equation}
    \bar l_1(a) = \int_0^\infty d\tilde q \, \tilde q^{2a} \frac{\partial_t R_k(\tilde q,0)}{Z_x k^2} \tilde P_k(\tilde q,0)^2 .
\end{equation}
Thus, $\tilde W_{3.k}(u)$ decouples from the other flow equations, as its contribution to the flow equation of $\tilde \Delta_k(u)$ vanishes. This also explains the explicit appearance of the parameter $\tilde t_k=\pi K_k \tilde T_k$ in the finite-temperature flow equations. 

On the other hand, one can show that the asymptotic behavior of $\bar M^\tau_{n,m}(a)$ is given by
\begin{align} \label{asympt_Mt}
    \bar M^\tau_{n,m}(a) =& \frac{1}{2}\biggl(\frac{-1}{2\tilde\Omega^2}\biggr)^m \bar l_{n+m}(a) 
    + \frac{m}{4}\biggl(\frac{-1}{2\tilde\Omega^2}\biggr)^{m+1} \nonumber\\ & \times \bar l_{n+m-1}(a) + \mathcal{O}\bigl(\tilde\Omega^{-2(m+2)} \bigr) . 
\end{align}
This behavior ensures that Eq.~\eqref{W3eq} takes a simple form in the limit $\tilde t_k\gg 1$, 
\begin{equation}
\partial_t \tilde{W}_{3,k}(u) =  - \frac{\tilde t_k}{\pi} \bar l_1(0)   \tilde{W}_{3,k}''(u) .
\end{equation}
Since $\bar l_1(0)>0$, the function $\tilde W_{3,k}$ is irrelevant in this limit.

\section{Beyond the derivative expansion}
\label{app_beyondDE} 

The singular behavior of the Luttinger parameter $K_k$ at $k_c$ suggests that the two-point vertex $\Gamma_{1,k}^{(2)}(q,i\omega_n)$ becomes a nonanalytic function of $\omega_n$ when $k<k_c$. If this is the case, we must go beyond the derivative expansion. Here, we discuss two possible directions. 

\subsection{LPA$''$ ansatz}

The ansatz~\eqref{ansatz1} resembles the LPA$'$~\cite{Dupuis_review}, since there is no field dependence in the derivative terms. As such a dependence is forbidden by the STS, the LPA$'$ is the more general ansatz to second order in the derivative expansion. As a consequence, generalizing the ansatz~(\ref{ansatz1},\ref{ansatz2}) to a form akin to the LPA$''$~\cite{Rose2018} constitutes a natural improvement. We therefore consider the following ansatz for $\Gamma_{1,k}$,
\begin{equation}
	\Gamma_{1,k}[\phi_a]=\int_{x,\tau} \biggl\{ \frac{Z_{x,k}}{2}(\dx \phi_a)^2+\frac{1}{2}\phi_a \Sigma_{1,k}(-\dtau)\phi_a \biggr\} ,
	\label{LPA''}
\end{equation}
with the initial condition $\Sigma_{1,\Lambda}(-\dtau)=-\frac{1}{\pi v K}\partial^2_\tau$. The regulator function must then be modified by replacing $\omega_n^2/v_k^2$ with $\Sigma_{1,k}(i\omega_n)/Z_x$. This implies that $\partial_t\tilde\Sigma_{1,k}(i\tilde\omega)$ enters the threshold functions, and its dimensionless form $\tilde\Sigma_{1,k}(i\tilde\omega)=\Sigma_{1,k}(i\omega)/Z_xk^2$  (with $\tilde\omega$ being defined through a proper prescription) obeys an equation of the type
\begin{equation}
  \partial_t\tilde\Sigma_{1,k}(i\tilde\omega) = f+ g \partial_t\tilde\Sigma_{1,k} (i\tilde\omega) ,
\label{eqSig}
\end{equation}
where $f$ and $g$ are $k$-independent functions of $\tilde\omega$, $\tilde\Sigma_{1,k}(i\tilde\omega)$, $\tilde\Delta_k''(0)$, $\tilde W_{1,k}''(0)$, and $\tilde W_{3,k}''(0)$. Equation~\eqref{eqSig} does not have a solution if $g=1$ for some $\tilde\omega$ at a given RG time. The numerical solution of the flow equations shows that this occurs at a small RG time, which points to the inadequacy of the ansatz~\eqref{LPA''}. It should be noted that the LPA$''$ considered here does not account for possible nonanalyticities in frequency of the two-replica term $\Gamma_{2,k}[\phi_a,\phi_b]$. This motivates an exploration of a more powerful framework to address singular frequency dependence, as described in the next section.

\subsection{BMW approximation}

The Blaizot--Méndez-Galain--Wschebor (BMW) approximation scheme \cite{Blaizot2006,Benitez2009,Benitez2012} is specifically designed to capture, albeit approximately, the full momentum dependence of the one-particle-irreducible (1PI) vertices obtained from $\Gamma_k$. It has been applied to the $\varphi^4$ theory~\cite{Benitez2009,Benitez2012,Rose2016} and to interacting boson systems \cite{Dupuis2009,Rancon2011,Rancon2014}. We recall here its principle for a simple, one-dimensional, scalar theory, so as to keep this section self-contained. Starting from the Wetterich equation, one can obtain flow equations for the 1PI vertices by functionally differentiating with respect to the field $\phi$. Fourier transforming, working at constant field, and considering the case of $\Gamma_k^{(2)}(p;\phi)$ for concreteness, one obtains a flow equation whose right-hand side involves the vertices $\Gamma_k^{(3)}(p,q,-p-q;\phi)$ and $\Gamma_k^{(4)}(p,-p,q,-q;\phi)$. The regulator function $R_k(q)$ restricts the momentum integrals to the range $|q|\lesssim k$, which in turn allows us to set $q=0$ in the three- and four-point vertices. Using the exact relations
\begin{equation}
\begin{split} 
	\Gamma_k^{(3)}(p,0,-p;\phi) &= \partial_\phi\Gamma_k^{(2)}(p;\phi) , \\
	\Gamma_k^{(4)}(p,-p,0,0;\phi) &= \partial_\phi^2 \Gamma_k^{(2)}(p;\phi) , 
\end{split}
\end{equation}
one obtains a closed flow equation for $\Gamma^{(2)}(p;\phi)$, which can then be solved numerically. 

This procedure can be generalized to the replicated effective action $\Gamma_k[\{\phi_a\}]$ and its cumulant expansion, which leads to closed flow equations for the vertices $\Gamma_{1,k}^{(2)}(p)$, $\Gamma_{2,k}^{(20)}(p;u)$, and $\Gamma_{2,k}^{(11)}(p;u)$. The fact that $\Gamma_{1,k}^{(2)}$ does not depend on the fields, and that $\Gamma_{2,k}^{(20)}$ and $\Gamma_{2,k}^{(11)}$ depend only on $u=\phi_a-\phi_b$, is preserved under the flow. We simplify the numerical integration by using an ansatz of the form
\begin{equation}
\begin{split} 
	\Gamma_{1,k}^{(2)}(p) & = Z_{x,k} q^2 + \Sigma_{1,k}(i\omega) , \\
	\Gamma_{2,k}^{(11)}(p;u) & = \Delta_k(u) + W_{1,k}(u) q^2 , \\
	\Gamma_{2,k}^{(20)}(p;u) & = -\Delta_k(u) + W_{2,k}(u) q^2 + \Sigma_{2,k}(i\omega,u) ,
\end{split}
\end{equation}
which amounts to promoting the term corresponding to $W_{3,k}$ in the derivative expansion to a full function of both $u$ and $\omega$. We use a second-order derivative expansion for the spatial dependence, and the two-point vertex $\Gamma_{1,k}^{(2)}(p)$ includes a frequency-dependent self-energy, as in the preceding section. As expected, the BMW flow equations do not exactly reproduce the second order of the derivative expansion in the small-frequency limit. However, a closer examination shows that, while all terms of the derivative expansion flow are present, one of them acquires the opposite sign because of a contribution ignored in the BMW approximation. This results in a pathological behavior of the flow, with a dynamical critical exponent $z_k$ that becomes negative.


%


\begin{thebibliography}{88}
\bibitem [{no1()}]{no1}%
  \BibitemOpen
  \bibinfo {note} {Let us introduce the function $Y(u)$ such that $Y'(u)=W_4(u)$. Then $\int_{\tau} \dtau Y(\phi_a-\phi_b)=\int_\tau W_4(\phi_a-\phi_b) \dtau\phi_a$.}\BibitemShut {Stop}%
\bibitem [{no2()}]{no2}%
  \BibitemOpen
  \bibinfo {note} {This can be seen from the fact that the long-wavelength part of the density field $\rho(x,\tau)=\rho_0-\frac{1}{\pi} \dx\varphi(x,\tau)$ is even under parity.}\BibitemShut {Stop}%
\end{thebibliography}

\end{document}